\documentclass[twocolumn,amsmath,amsfonts,amssymb,aps,prx,preprintnumbers,superscriptaddress]{revtex4-2}

\usepackage[utf8]{inputenc}
\usepackage[T1]{fontenc}
\usepackage{lmodern}
\usepackage{graphicx}
\usepackage{dcolumn}
\usepackage{bm}
\usepackage{textcomp}
\usepackage{float}

\usepackage[normalem]{ulem}
\usepackage{ifpdf}
\usepackage[squaren,Gray]{SIunits}

\usepackage{amsmath}

\ifpdf
\usepackage{epstopdf}
\usepackage[pdftex,unicode,pdfstartview={FitH},pdfborder={0 0 0}]{hyperref}
\usepackage{hypcap}
\else
\usepackage[hypertex]{hyperref}
\fi
\hypersetup{
	bookmarksnumbered = true,
	colorlinks = true, linkcolor = darkblue,
	citecolor = darkblue, filecolor = darkblue,
	menucolor = darkblue, urlcolor = black
}

\usepackage{color}
\definecolor{red}{rgb}{1,0,0}
\definecolor{blue}{rgb}{0,0,1}
\definecolor{black}{rgb}{0,0,0}

\usepackage{color}
\definecolor{darkblue}{rgb}{0,0,0.6}

\definecolor{red}{rgb}{1,0,0}

\definecolor{green}{rgb}{0,0.6,0}

\definecolor{grey}{rgb}{0.7,0.7,0.7}

\definecolor{orange}{rgb}{0.8,0.4,0}

\usepackage{titlesec}           
\titleformat{\subsection}
{\bfseries} 
{}          
{0.0cm}     
{}          
[]          

\begin{document}

\title{Cavity-mediated exciton hopping in a dielectrically engineered polariton system}

\author{Lukas Husel}

\def\LMU{Fakult\"at f\"ur Physik, Munich Quantum Center, and Center for NanoScience (CeNS), Ludwig-Maximilians-Universit\"at M\"unchen, Geschwister-Scholl-Platz~1, D-80539 M\"unchen, Germany}
\affiliation{\LMU}

\author{Farsane Tabataba-Vakili}
\affiliation{\LMU}
\affiliation{Munich Center for Quantum Science and Technology (MCQST), Schellingstr.~4, D-80799 M\"unchen, Germany}  
\affiliation{Institute of Condensed Matter Physics, Technische Universität Braunschweig, 38106 Braunschweig, Germany }

\author{Johannes Scherzer}
\affiliation{\LMU}

\author{Lukas Krelle}
\affiliation{\LMU}
\def\DAR{Present affiliation: Institute for Condensed Matter Physics, TU Darmstadt, Hochschulstr.~6-8, D-64289 Darmstadt, Germany}
\affiliation{\DAR}

\author{Ismail Bilgin}
\affiliation{\LMU}

\author{Samarth Vadia}
\affiliation{\LMU}

\author{Kenji Watanabe}
\affiliation{Research Center for Electronic and Optical Materials, National Institute for Materials Science, 1-1 Namiki, Tsukuba 305-0044, Japan}

\author{Takashi Taniguchi}
\affiliation{Research Center for Materials Nanoarchitectonics, National Institute for Materials Science, 1-1 Namiki, Tsukuba 305-0044, Japan}

\author{Iacopo Carusotto}
\affiliation{Pitaevskii BEC Center, INO-CNR and Dipartimento di Fisica, Universita di Trento, via Sommarive 14, I-38123 Trento, Italy}  

\author{Alexander H\"ogele}
\affiliation{\LMU}
\affiliation{Munich Center for Quantum Science and Technology (MCQST), Schellingstr.~4, D-80799 M\"unchen, Germany}

\date{\today}

\begin{abstract}

Exciton-polaritons -- coherently hybridized states of excitons and photons -- are instrumental for solid-state nonlinear optics and quantum simulations. To enable engineered polariton energy landscapes and interactions, local control over the particle-like states can be achieved by tuning the properties of the exciton constituent. Monolayer transition metal dichalcogenides stand out in this respect, as they readily allow for a deterministic, flexible and scalable control of excitons, and thus of hybrid exciton-polaritons, via environmental dielectric engineering. Here, we demonstrate the realization of mesoscopic exciton-polariton domains in a structured dielectric exciton environment, and establish an effective long-range exciton hopping in the dispersive regime of cavity-coupling. Our results represent a crucial step toward interacting polaritonic networks and quantum simulations in exciton-polariton lattices based on dielectrically tailored two-dimensional semiconductors.
\end{abstract}

\maketitle

Coherently hybridized states of cavity photons and semiconductor excitons manifest as exciton-polaritons and underpin phenomena of non-equilibrium quantum many-body physics~\cite{Deng2010a, Carusotto2013} and quantum simulations~\cite{Basov2021} in conventional quantum well microcavity structures. In two-dimensional semiconductors and related van der Waals heterostructures, large exciton binding energies and oscillator strengths enable robust polariton formation and condensation at elevated temperatures \cite{Schneider2018,Anton-Solanas2021} with unique features provided by spin-valley locking \cite{Qiu2019,Lundt2019}, novel quasiparticles with enhanced nonlinearities upon doping \cite{Tan2020}, or moir\'e exciton-polaritons in heterostructures \cite{Zhang2021,Scherzer2024}. These properties establish semiconductor monolayers and heterostructures as prime candidates for studies of strong light-matter coupling in the solid-state. 
 
Control over the polariton degrees of freedom via the excitonic fraction is the key to recent developments in device design~\cite{Schneider2016} and has enabled demonstrations of topological insulators~\cite{Klembt2018} or Kardar-Parisi-Zhang universality~\cite{Fontaine2022} in conventional semiconductor quantum wells. Within this framework, excitons in monolayer transition metal dichalcogenides (TMDs) with high sensitivity to their dielectric environment provide means of additional engineering, taking advantage of sizable changes in the band gap and binding energy induced by proximal dielectrics~\cite{Borghardt2017, Raja2017, Peimyoo2020, BenMhenni2025}. As such, local disorder in TMD devices facilitates trapping of polaritons with enhanced coherence~\cite{Anton-Solanas2021,Wurdack2021}, while complementary approaches have demonstrated polariton trapping by additional monolayers~\cite{Rupprecht2020,Wurdack2022}, laser-induced disorder~\cite{Li2023} and local strain~\cite{Zhao2021}.
Despite this progress, deterministic, scalable and tunable control over local exciton-polariton energies, potentials and couplings remains challenging.

\begin{figure*}[t]
\centering \includegraphics[scale=1.05]{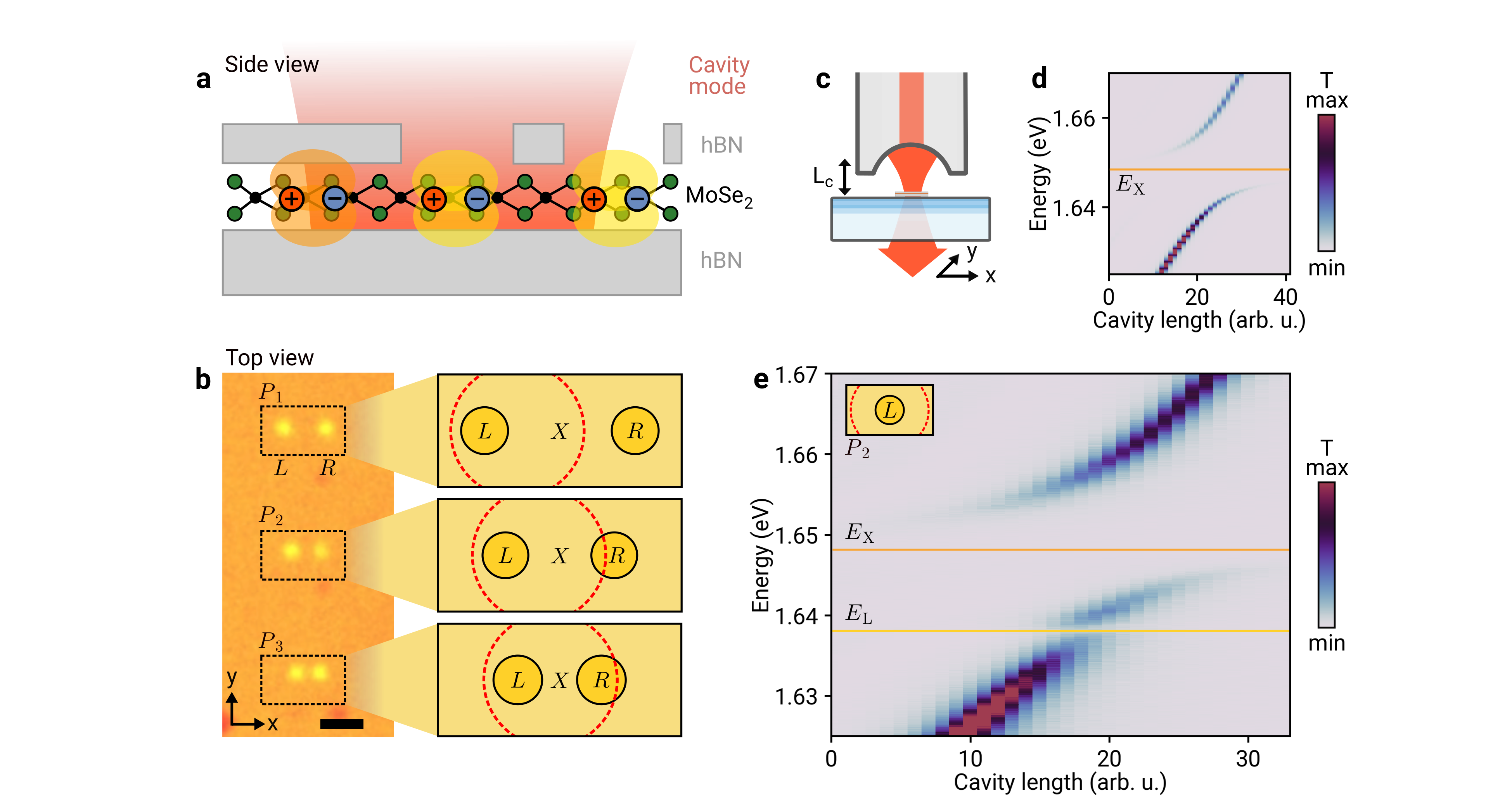}
\caption{\textbf{Dielectric engineering of polariton domains.} \textbf{a}, Schematics of a MoSe$_{2}$ monolayer encapsulated by planar bottom and patterned top hBN layers with spatially distinct regions of exciton-polaritons formed by strong coupling to the mode of an open cavity. \textbf{b}, Left panel: optical micrograph of the van der Waals heterostructure, with pairs of holes in the top hBN layer P$_1$, P$_2$ and P$_3$ visible as yellow circles of nominally identical left ($L$) and right ($R$) disks with variable separation, surrounded by fully encapsulated monolayer. The scale bar is $2~\mu$m. Right panel: The red dashed circle shows the waist diameter of the cavity mode on the scale of the diameters and distances of disk-shaped domains surrounded by regions of monolayer excitons ($X$) in unpatterned hBN. \textbf{c}, Schematic of the cryogenic fiber-based open microcavity, with piezoelectrically actuated lateral translation in the $x - y$ plane and cavity length ($L_{\mathrm C}$) tuning along $z$. \textbf{d}, Cavity transmission as a function of the cavity length tuned via the $z$-piezo voltage, recorded away from structured hBN. The exciton energy $E_{\mathrm X}$ is indicated by the horizontal solid line. \textbf{e}, Same but with the cavity mode positioned near the center of a single exciton disk of etched pair P$_2$ as illustrated in the inset. The solid horizontal lines indicate the energy of monolayer and disk-localized excitons $E_{\mathrm X}$ and $E_{\mathrm L}$, respectively.}
\label{fig1}
\end{figure*}

Here, we demonstrate control of local exciton-polariton energies via environmental dielectric engineering and spectral cavity detuning, and show how the spatially confined cavity mode can be used to mediate hopping-like polariton coupling. To define local exciton sites, we use a nanopatterned encapsulation layer of hexagonal boron nitride (hBN), which consolidates disk-shaped exciton domains with energies distinct from excitons in the surrounding monolayer regions with unpatterned hBN. Strong coupling of one or multiple exciton domains to the optical mode of a fiber-based microcavity results in distinct polariton states, with properties determined by both the locally engineered dielectric environment and the actively controlled spectral cavity-coupling. The resulting modulation in the lower polariton energy landscape is evidenced in transmission spectroscopy of the the strongly-coupled fiber cavity. Moreover, we demonstrate how the regime of dispersive cavity coupling~\cite{Blais2004} mediates hopping between distant domains of excitons weakly dressed by cavity photons as a premise to quantum simulation architectures, complementing already established for superconducting qubits~\cite{Majer2007}, ultracold atoms~\cite{Ritsch2013}, solid-state quantum emitters~\cite{Evans2018} and optomechanical systems~\cite{Vijayan2024}.

\begin{figure*}[t]
\centering \includegraphics[scale=1.05]{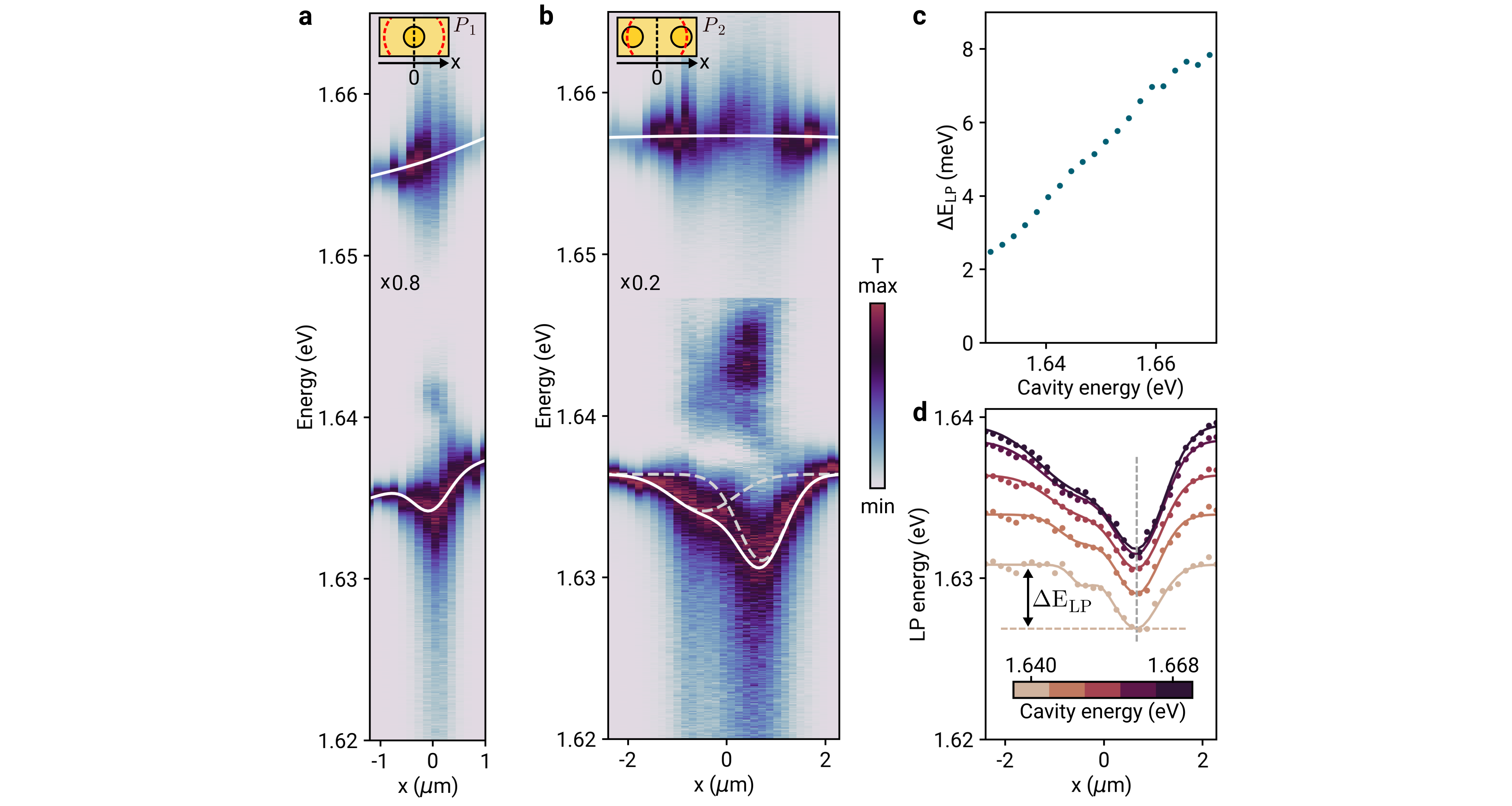}
\caption{\textbf{Dielectrically engineered polariton disks with cavity-controlled local energy modulation.} \textbf{a}, Cavity transmission for a fixed cavity energy and lateral cavity mode translation across the right exciton domain of pair P$_1$, as illustrated in the inset. The cavity energy corresponds to an exciton-cavity detuning $\Delta_X = -5$~meV at position $x=-1.2~\mu$m. The white solid lines show best fits of polynomial and Gaussian profiles to maximum transmission of the upper and lower polariton branches, respectively. \textbf{b}, Same but for the domain pair P$_2$, recorded at similar detuning of $\Delta_X = -7$~meV at $x=-2.2~\mu$m ($x=0~\mu$m corresponds to the cavity mode positioned in between the two exciton disks, as illustrated in the inset). The dashed lines are the individual Gaussian contributions to the full lower polariton transmission profile shown by the solid line. The spectra at each position were normalized to the maximum transmission of the lower polariton branch and rescaled for the upper polariton branch by a factor of $0.8$ and $0.2$ as indicated in the respective sub-panels. \textbf{c}, Lower polariton energy shift $\Delta E_{\mathrm{LP}}$ as a function of the cavity energy, defined as illustrated in \textbf{d} for the right polariton disk of P$_2$. \textbf{d}, Energy profiles of the lower polaritons (corresponding to the transmission maxima in \textbf{b}) as a function of the lateral cavity displacement for five cavity energies indicated by the color-bar. The solid lines show best fits of two Gaussians to the polariton energy profiles at different cavity energies.}
\label{fig2}
\end{figure*}

Our approach, illustrated schematically in Fig.~\ref{fig1}a, is based on a van der Waals heterostack with a monolayer of molybdenum diselenide (MoSe$_2$) encapsulated by a planar bottom hBN layer and a patterned top hBN layer with through-holes defined by reactive-ion etching (see the Methods section for details). The resulting modification of the dielectric environment on the scale of a few hundred of nanometers affects both the monolayer band gap and the exciton binding energy~\cite{Borghardt2017,Raja2017,BenMhenni2025}, yielding a spatially modified exciton resonance energy which in turn determines local properties of exciton-polaritons upon strong coupling to an optical microcavity with photonic modes confined in all spatial directions. The left panel in Fig.~\ref{fig1}b shows an optical micrograph of the corresponding sample, with two holes in the top hBN layer of same diameter and variable distance as pairs of yellow circles. The right panel illustrates schematically the resulting exciton landscape: left ($L$) and right ($R$) disk-shaped exciton domains at each etch site are surrounded by monolayer excitons ($X$) in the unpatterned area. Due to fabrication imperfections such as interfacial bubbles and unintentional strain, the exciton energies $E_{\mathrm{L}}$ and $E_{\mathrm{R}}$ differ between the two nominally identical domains, and both differ from the surrounding exciton energy $E_{\mathrm{X}}$ by virtue of different dielectric environments. 

In the following, we focus on three etch-site pairs P$_1$, P$_2$ and P$_3$ in Fig.~\ref{fig1}b with $2.0$, $1.4$ and $1.1~\mu$m distances between the centers of the left and right holes with identical diameters of $0.6~\mu$m. The finite extent of the fundamental Gaussian cavity mode with a waist of $\sim 1~\mu$m (shown by the red dashed circle in the schematics of Fig.~\ref{fig1}b) and the tunability of the resonance energy as well as the lateral mode position of our open cavity (as indicated in Fig.~\ref{fig1}c) allows us to study different limits of polaritons in strong light-matter coupling. First, by placing the cavity mode over the left site of the pair P$_1$ or P$_2$ that is sufficiently distant from its right counterpart, we study the local formation of polaritons in the left disk as well as their coupling to the surrounding exciton-polariton continuum (top panel of the schematics in Fig.~\ref{fig1}b). In the second setting (central panel in Fig.~\ref{fig1}b), the cavity mode creates and samples both left and right polariton disks, yet at a distance too large for intersite coupling. The third configuration (bottom panel in Fig.~\ref{fig1}b), finally, is used to demonstrate effective cavity-mediated coupling between the left and right sites of polariton pairs.   

We begin by calibrating the coupling strength between the exciton domains and our tunable microcavity in a closed-cycle cryostat with a base temperature of $4.3$~K~\cite{Vadia2021} according to the schematics in Fig.~\ref{fig1}c (see the Methods section for details on the cavity setup). The nanopatterned heterostack is placed on a macroscopic planar mirror, whose vertical separation $L_{\mathrm C}$ from the microscopic mirror of the micromachined fiber facet is controlled by piezoelectric actuators, which also allow for lateral displacement of the sample with respect to the cavity mode. As such, the spectral resonance condition between the exciton and cavity energy, $E_{\mathrm X}$ and $E_{\mathrm C}$ is tunable via the cavity length and exhibits a clear signature of strong-coupling in the cavity transmission of Fig.~\ref{fig1}d on a region away from structured hBN. The avoided crossing is a hallmark of polariton formation, with light-matter coupling strength $g_{\mathrm X} = 9.6$~meV at longitudinal cavity mode order $q=6$ as determined from the dissipative model analysis (see Supplementary Note II for details). This coupling strength is characteristic for cavity-coupling of monolayer excitons~\cite{Vadia2021,Tan2023}, and places the system together with polariton linewidths of $\sim 2.5$~meV and the cavity linewidth $\kappa \simeq 1.5$~meV (limited by residual vibrational fluctuations in the cavity length \cite{Vadia2021}) in the regime of strong light-matter coupling~\cite{Savona1995}.

\begin{figure*}[t]
    \centering \includegraphics[scale=1.05]{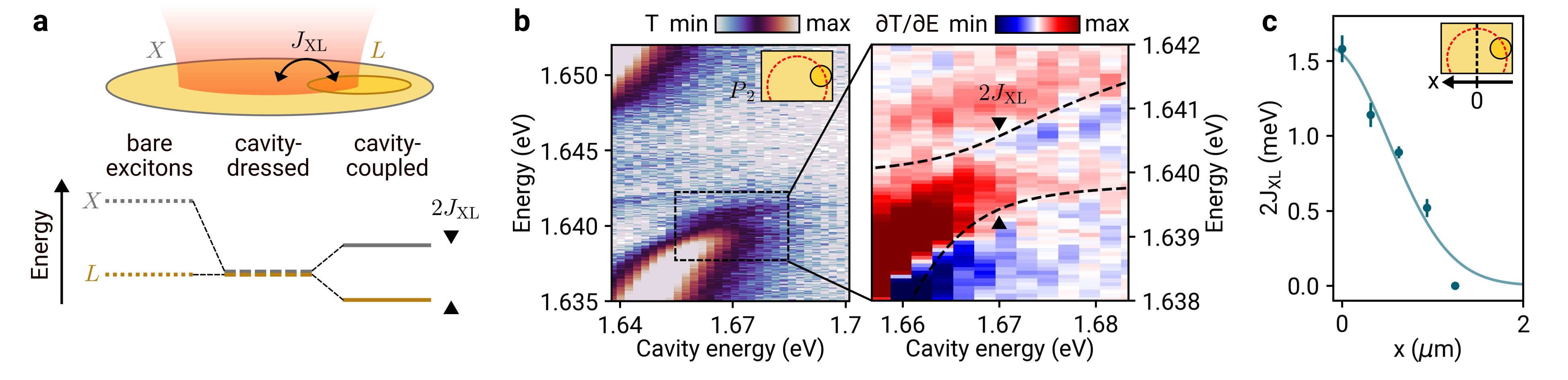}
    \caption{\textbf{Cavity-mediated long-range exciton hopping.} \textbf{a}, Top: Schematic of excitons $L$ and $X$, coupled via an effective cavity-mediated hopping of strength $J_{\mathrm{XL}}$. Bottom left: exciton energies in the absence of a cavity mode. Bottom center: exciton energies tuned into resonance due to dispersive cavity-dressing of each individual exciton state. Bottom right: Exciton-like system eigenstates as observed in experiment, split due to the effective interaction $J_{\mathrm{XL}}$. \textbf{b}, Left panel: cavity transmission as a function of cavity energy, with the cavity mode positioned at the edge of the domain $L$ as illustrated in the inset. Right panel: derivative of cavity transmission with respect to energy, computed for data in the dashed rectangle in the left panel. The dashed lines are the eigenstates of the effective system Hamiltonian, Eq~\ref{Eq:Ht}. \textbf{c}, $J_{\mathrm{XL}}$ as a function of the cavity mode position, which is moved away from the etch site center ($x = 0~\mu$m corresponds to the mode position illustrated in the inset).}
    \label{fig3}
\end{figure*}

We observe a strong modification of the characteristic exciton-polariton splitting as we position the cavity over the left hole-etched site of P$_2$, with cavity transmission shown in Fig.~\ref{fig1}e. As a function of the cavity length detuning, we observe an additional polariton branch related to the disk-localized exciton fraction with energy $E_{L}$, redshifted from $E_{\mathrm X}$ by $10$~meV due to effectively reduced screening below the hole in hBN. As the area of the local exciton domain is smaller than the cavity spot, the corresponding light-matter coupling strength is reduced to $g_{L} = 2.65 \pm 0.04$~meV as compared to the coupling of spatially unconfined monolayer excitons in regions with both-sided hBN encapsulation (see Supplementary Note III for details). This scaling of the light-matter coupling with the spatial extent of exciton-confining domains with redshifted transition energy indicates the formation of local polariton disks.

The creation of exciton domains results in local shifts of the polariton energy, which we map out by spatial and spectral cavity tuning with data shown in Fig.~\ref{fig2}. Displacing the cavity laterally at a constant cavity length across one domain of the pair P$_1$, the cavity transmission leads to the evolution of the upper and lower polariton branches as in Fig.~\ref{fig2}a. In agreement with the data in Fig.~\ref{fig1}e, the middle polariton branch is visible at an energy of $\sim$ 1.640~eV. The intensity of this feature is brightened as a result of spectral overlap with higher order transverse cavity modes (see Suplementary Note II for details). Crucially, the lower polariton branch exhibits an energetic minimum at the center of the etch site. We emphasize that this redshift of $\sim 2$~meV would correspond to an attractive polariton potential for spatially extended polaritons in two-dimensional cavities~\cite{Rupprecht2021}, thereby providing lateral confinement for exciton-polaritons. The superimposed near-linear energy gradient, also evidenced in the upper polariton branch, stems from the unintentional spatial gradient of the exciton energy $E_X$ around this site. 

The effect of the nanostructured dielectric environment on the polariton energy is even richer in the pair P$_2$ with transmission data in Fig.~\ref{fig2}b. Upon spatial displacement across the left and right disk of the pair for a constant cavity energy, we observe in Fig.~\ref{fig2}b two local energy minima for the lower polariton branch, indicated by the two Gaussian contributions (dashed lines) to the full transmission profile (solid line). The difference in the left and right local energy shift is related to differences in the disks from imperfect fabrication. Remarkably, the energy shift $\Delta E_{\mathrm{LP}}$ is tunable via the cavity energy, as evident from the dependence of the lower polariton energies on the cavity energy shown for five discrete values in Fig.~\ref{fig2}d. A systematic study of the lower polariton energy shift, shown in Fig.~\ref{fig2}c, reveals a monotonous increase in $\Delta E_{\mathrm{LP}}$ with increasing cavity energy, spanning a range of several meV.

The dependence of $\Delta E_{\mathrm{LP}}$ on the cavity resonance condition can be understood by noting that for large cavity energy, the lowest polariton energy in the system is dictated by the lowest-energy exciton. By virtue of dielectric engineering, the exciton energy at the etch-site center is reduced from its monolayer value in fully hBN-encapsulated regions. Thus, in the limit of large cavity energies, the local energy shift for the lower polariton corresponds to the energy difference imprinted by different dielectric environments. 

\begin{figure*}[t]
\centering \includegraphics[scale=1.05]{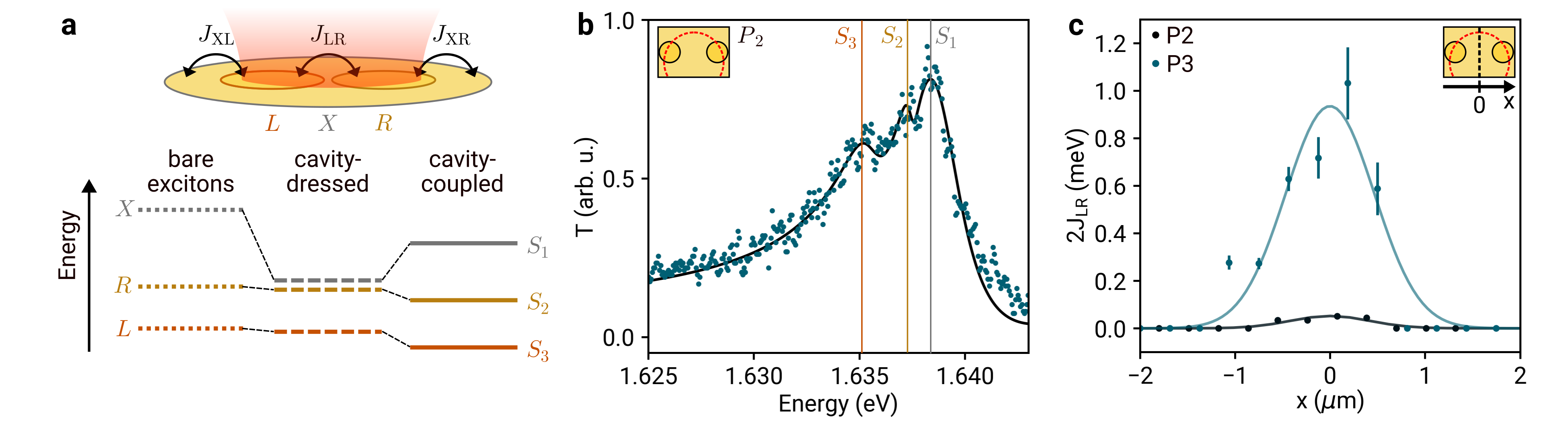} 
\caption{\textbf{Effective inter-site and site-to-surrounding hopping.} \textbf{a}, Coupling $J_{\mathrm{LR}}$ between excitons of left and right domain $L$ and $R$, mediated by the cavity in addition to the respective couplings $J_{\mathrm{XL}}$ and $J_{\mathrm{XR}}$ with the surrounding exciton reservoir $X$. \textbf{b}, Cavity transmission spectrum (dark green data) for a cavity energy of $1.659$~eV and mode position as illustrated in the inset with best fit (black solid line) according to the dissipative model analysis. The vertical lines indicate the eigenenergies of the coupled system. \textbf{c}, Coupling strength $J_{\mathrm{LR}}$ at cavity energy 1.670~eV for the etch site pairs P$_2$ and P$_3$ (black and dark green data, respectively, with error bars of one standard deviation and Gaussian fits as solid lines) as the cavity is moved across the etch site pair, as obtained from exciton light-matter coupling strengths and resonance energies ($x = 0~\mu$m corresponds to the cavity mode centered between both sites).}
\label{fig4}
\end{figure*}

The second key feature of our system is the ability to establish site-to-surrounding and site-to-site coupling in the dispersive cavity regime, mediating an effective long-range hopping. From a theoretical perspective, this coupling follows from the time-independent Hamiltonian for multiple cavity-coupled exciton domains $i$ with light-matter coupling strengths $g_i$, energies $E_i$ and detunings $\Delta_i = E_i - E_C$ from the cavity energy $E_C$. Expansion to second order in $g_i$/$\Delta_i$ via a Schrieffer-Wolff transformation~\cite{Blais2004, Majer2007} yields:
\begin{multline}
	\tilde{H} \simeq \left(E_{C} - \sum_i \frac{g_i^2}{\Delta_i}\right)a^{\dagger}a + \sum_i \left(E_{i} +\frac{g_i^2}{\Delta_i}\right) b_i^{\dagger}b_i\\  + \sum_i \sum_{j\neq i} \frac{g_i g_j}{2\Delta_i}\left(b_i^{\dagger}b_j+b_j^{\dagger}b_i\right), 
	\label{Eq:Ht}
\end{multline}
with the respective bosonic annihilation operators for cavity photons and excitons, $a$ and $b_i$. The result is a system described by multiple exciton resonances which are weakly dressed by cavity photons, evidenced by the diagonal terms in the first line of Eq.~\ref{Eq:Ht}. In addition, excitons associated with different resonances $i$ and $j$ are coupled via an effective beam-splitter type coupling of strength $J_{ij} =g_i g_j \left(\Delta_i^{-1} + \Delta_j^{-1}\right)/2$ as described by the second line of Eq.~\ref{Eq:Ht}, mediated by dispersive exchange of cavity photons. 

The resulting effective coupling between disk-localized excitons and their surrounding monolayer excitons is sizable when the cavity mode is positioned close to the edge of a single etch site. This effect is shown conceptually in Fig.~\ref{fig3}a: since the light-matter coupling strength of $X$ excitons greatly exceeds that of $R$ and $L$ domains, the dispersive shift of magnitude $g_i^2/\Delta_i$ (see Eq.~\ref{Eq:Ht}) can be used to tune the difference in the respective energies of the bare excitons (left panel) into resonance (central panel). Here, dispersive coupling induces an effective splitting given by $2 J_{\mathrm{LX}}$ (right panel). The corresponding experiment on the left disk in pair P$_2$ shows in Fig.~\ref{fig3}b the avoided crossing due to dispersive cavity-dressing and coupling, with a maximum coupling strength of $2 J_{\mathrm{LX}}=1.5$~meV for optimal conditions. In this regime, the eigenstates of the system derive from the Hamiltonian of Eq.~\ref{Eq:Ht}, with energies shown by the black dashed lines in the right panel of Fig.~\ref{fig3}b (see Supplementary Note II for details). Unavoidably, the coupling vanishes as the cavity is moved away from the etch-site, as confirmed in Fig.~\ref{fig3}c.

Finally, we demonstrate effective long-range coupling mediated by the cavity between excitons of two distant sites, as shown conceptually in Fig.~\ref{fig4}a: we use cavity-dressing to tune the energy of surrounding excitons $X$ close to the resonances of the disks $L$ and $R$. With dispersive coupling, we obtain hybrid eigenstates S$_1$, S$_2$ and S$_3$ of the coupled system, with energies defined by effective interactions $J_{\mathrm{XL}}$, $J_{\mathrm{XR}}$ and $J_{\mathrm{LR}}$ among all three constituents. The corresponding spectral signature is shown in Fig.~\ref{fig4}b, with maximal intersite coupling strength $2 J_{\mathrm{LR}}$ of $\sim 0.1$~meV and $1$~meV obtained for the pairs P$_2$ and P$_3$, respectively. In the framework of dispersive cavity-coupling, these exciton resonance energies differ from the bare eigenstates of the system as a direct result of the cavity-induced interaction. This observation is enabled by two features of our system: the energy difference of $L$ and $R$ excitons, a result of our choice of cavity mode position and sample inhomogeneities, renders all three eigenstates optically bright, while the tunability of our fiber cavity provides access to the relevant cavity energy. Lateral displacement of the cavity mode results in the reduction of coupling strength, as expected from Eq.~\ref{Eq:Ht} and evidenced in Fig.~\ref{fig4}c.

To conclude, we have demonstrated the realization of dielectrically engineered exciton-polariton domains with cavity-mediated long-range exciton hopping in the dispersive coupling regime. We anticipate that our approach will enable polariton confinement in two-dimensional cavities to leverage engineered geometries of polaritonic lattices. In combination with active gate control of exciton confinement~\cite{Thureja2022,Hu2024} and extended to Rydberg excitons~\cite{Gu2021} or hybrid moir\'e excitons~\cite{Gu2024,Polovnikov2024} with enhanced nonlinearities, our technique represents a promising approach for advancing quantum simulations in exciton-polariton lattices and circuits~\cite{Boulier2020}.\\

\noindent \textbf{METHODS}

\noindent \textbf{Sample fabrication:} Our device is based on a monolayer of MoSe$_2$ grown by chemical vapor deposition and encapsulated in flakes of hBN obtained by mechanical exfoliation. The bottom hBN thickness was $87$~nm (as determined with atomic force microscopy), placing the TMD monolayer close to an antinode of the intracavity field. A polymer mask with a through-hole pattern was fabricated with electron beam lithography on the top hBN flake with a thickness of $44$~nm. 4\% PMMA 950K dissolved in anisole was used as resist, with typical electron beam doses and acceleration voltages of $50~\mu$C/cm$^2$ and $20$~kV, respectively. Through-holes were etched using inductively coupled plasma-based reactive ion (ICP-REI) etching. We used a plasma of Ar and SF$_6$, which were injected at flowrates of $5$ and $10$~sccm, respectively, as the chamber pressure was kept at $10$~mTorr. ICP and RF powers were set to $70$ and $6$~W, respectively, with a resulting etch rate of $0.6$~nm/s. After etching and resist removal, the hBN flake was cleaned using oxygen plasma. Supplementary Fig.~6 shows a scanning electron micrograph of a flake with fabricated through-holes. The complete van der Waals stack was assembled using dry transfer using stamps based on polycaprolactone (PCL) polymer~\cite{Son2020, Shin2021} to ensure sufficient adhesion to the processed top hBN flake, since transfer attempts based on the widely used polycarbonate were unsuccessful. After stack assembly and release on the cavity mirror, PCL residues were removed using tetrahydrofuran. We note that our fabrication technique also forms the basis for our recent demonstration of a plasmonic metasurface~\cite{Tabataba-Vakili2024}. \\

\noindent \textbf{Cryogenic cavity system:} The fiber-based cavity system, described in detail in~\cite{Vadia2021}, was operated in a closed-cycle croystat (attocube attoDRY800) at a base temperature of $4.3$~K. The cavity was mounted on a passive vibration isolation system to suppress changes in cavity length induced by mechanical vibrations of the cryostat cold plate. A Gaussian-shaped indentation with radius of curvature of $\sim 14~\mu$m on the fiber tip served as concave cavity mirror. Fiber and planar mirror had identical dielectric coatings, forming highly reflective distributed Bragg reflectors. The cavity transmission was measured with a white-light source (NKT SuperK) in a spectral bandwidth of $10$~nm. The cavity mode transmitted through the planar mirror was collimated using an aspheric lens, before being coupled into a single mode fiber, dispersed in a grating spectrometer and detected with a CCD camera. The cavity also featured higher order Hermite-Gaussian modes~\cite{Mader2022}, whose influence on the transmission measurements is discussed in Supplementary Note II. 

The cavity was operated at longitudinal mode order $q=6$ as the lowest accessible without physical contact between the fiber and the planar mirror. At this mode order, the measured mode waist was $1~\mu$m, with a 15\% in-plane anisotropy. Typical integration times for transmission measurements were chosen between one and two seconds in order to average over multiple cryostat compressor cycles. As a result, the measured cavity transmission profiles are broadened by mechanical vibrations. With the cavity mode positioned away from the monolayer, we determined a full-width at half-maximum cavity linewidth $\kappa = 1.52$~meV from the fit of a Voigt profile, with Lorentzian and Gaussian contributions of $1.15$ and $0.70$~meV stemming from broadening by mirror loss and mechanical vibrations, respectively.
\newpage
\noindent \textbf{ACKNOWLEDGEMENTS}  

\noindent We thank David Hunger for providing the cavity fiber, as well as Philipp Altpeter and Christian Obermayer for assistance in the clean-room. This research was funded by the European Research Council (ERC) under the Grant Agreement No.~772195 as well as the Deutsche Forschungsgemeinschaft (DFG, German Research Foundation) within the Priority Programme SPP~2244 2DMP and the Germany's Excellence Strategy EXC-2111-390814868 (MCQST). I.\,B. acknowledges support from the Alexander von Humboldt Foundation. L.\,H. and A.\,H. acknowledge funding by the Bavarian Hightech Agenda within the EQAP project. F.T.-V. acknowledges funding from the European Union’s Framework Programme for Research and Innovation Horizon Europe under the Marie Skłodowska-Curie Actions grant agreement no. 101058981. K.\,W. and T.\,T. acknowledge support from the JSPS KAKENHI (Grant Numbers 21H05233 and 23H02052), the CREST (JPMJCR24A5), JST and World Premier International Research Center Initiative (WPI), MEXT, Japan. I.C. acknowledges financial support from the Provincia Autonoma di Trento, from the Q@TN Initiative, and from the National Quantum Science and Technology Institute through the PNRR MUR Project under Grant PE0000023-NQSTI, co-funded by the European Union - NextGeneration EU. \\

\noindent \textbf{AUTHOR CONTRIBUTIONS} 

\noindent L.~H., F.~T-V. and L.~K. fabricated samples using monolayer crystals synthesized by I.~B. and high-quality hBN crystals provided by K.~W. and T.~T.. L.~H. performed the experiments in a cryo-cavity implemented by J.~S. and S.~V.. L.~H., I.~C. and A.~H. analyzed the data. L.~H. and A.~H. wrote the manuscript. All authors commented on the manuscript. \\

\noindent \textbf{CORRESPONDING AUTHORS} \\
\noindent lukas.husel@physik.lmu.de, alexander.hoegele@lmu.de\\

\noindent \textbf{DATA AVAILABILITY}
  
\noindent The data that support the findings of this study are available from the corresponding authors upon reasonable request. \\

\noindent \textbf{COMPETING INTERESTS}  

\noindent The authors declare no competing interests.


\begin{thebibliography}{10}
	\expandafter\ifx\csname url\endcsname\relax
	\def\url#1{\texttt{#1}}\fi
	\expandafter\ifx\csname urlprefix\endcsname\relax\def\urlprefix{URL }\fi
	\providecommand{\bibinfo}[2]{#2}
	\providecommand{\eprint}[2][]{\url{#2}}

	\bibitem{Deng2010a}
	\bibinfo{author}{Deng, H.}, \bibinfo{author}{Haug, H.} \&
	\bibinfo{author}{Yamamoto, Y.}
	\newblock \bibinfo{title}{Exciton-polariton {{Bose-Einstein}} condensation}.
	\newblock \emph{\bibinfo{journal}{Reviews of Modern Physics}}
	\textbf{\bibinfo{volume}{82}}, \bibinfo{pages}{1489--1537}
	(\bibinfo{year}{2010}).
	
	\bibitem{Carusotto2013}
	\bibinfo{author}{Carusotto, I.} \& \bibinfo{author}{Ciuti, C.}
	\newblock \bibinfo{title}{Quantum fluids of light}.
	\newblock \emph{\bibinfo{journal}{Reviews of Modern Physics}}
	\textbf{\bibinfo{volume}{85}}, \bibinfo{pages}{299--366}
	(\bibinfo{year}{2013}).
	
	\bibitem{Basov2021}
	\bibinfo{author}{Basov, D.~N.}, \bibinfo{author}{{Asenjo-Garcia}, A.},
	\bibinfo{author}{Schuck, P.~J.}, \bibinfo{author}{Zhu, X.} \&
	\bibinfo{author}{Rubio, A.}
	\newblock \bibinfo{title}{Polariton panorama}.
	\newblock \emph{\bibinfo{journal}{Nanophotonics}}
	\textbf{\bibinfo{volume}{10}}, \bibinfo{pages}{549--577}
	(\bibinfo{year}{2021}).
	
	\bibitem{Schneider2018}
	\bibinfo{author}{Schneider, C.}, \bibinfo{author}{Glazov, M.~M.},
	\bibinfo{author}{Korn, T.}, \bibinfo{author}{H{\"o}fling, S.} \&
	\bibinfo{author}{Urbaszek, B.}
	\newblock \bibinfo{title}{Two-dimensional semiconductors in the regime of
		strong light-matter coupling}.
	\newblock \emph{\bibinfo{journal}{Nature Communications}}
	\textbf{\bibinfo{volume}{9}}, \bibinfo{pages}{2695} (\bibinfo{year}{2018}).
	
	\bibitem{Anton-Solanas2021}
	\bibinfo{author}{{Anton-Solanas}, C.} \emph{et~al.}
	\newblock \bibinfo{title}{Bosonic condensation of exciton-polaritons in an
		atomically thin crystal}.
	\newblock \emph{\bibinfo{journal}{Nature Materials}}
	\textbf{\bibinfo{volume}{20}}, \bibinfo{pages}{1233--1239}
	(\bibinfo{year}{2021}).
	
	\bibitem{Qiu2019}
	\bibinfo{author}{Qiu, L.}, \bibinfo{author}{Chakraborty, C.},
	\bibinfo{author}{Dhara, S.} \& \bibinfo{author}{Vamivakas, A.~N.}
	\newblock \bibinfo{title}{Room-temperature valley coherence in a polaritonic
		system}.
	\newblock \emph{\bibinfo{journal}{Nature Communications}}
	\textbf{\bibinfo{volume}{10}}, \bibinfo{pages}{1513} (\bibinfo{year}{2019}).
	
	\bibitem{Lundt2019}
	\bibinfo{author}{Lundt, N.} \emph{et~al.}
	\newblock \bibinfo{title}{Optical valley {{Hall}} effect for highly
		valley-coherent exciton-polaritons in an atomically thin semiconductor}.
	\newblock \emph{\bibinfo{journal}{Nature Nanotechnology}}
	\textbf{\bibinfo{volume}{14}}, \bibinfo{pages}{770--775}
	(\bibinfo{year}{2019}).
	
	\bibitem{Tan2020}
	\bibinfo{author}{Tan, L.~B.} \emph{et~al.}
	\newblock \bibinfo{title}{Interacting {{Polaron-Polaritons}}}.
	\newblock \emph{\bibinfo{journal}{Physical Review X}}
	\textbf{\bibinfo{volume}{10}}, \bibinfo{pages}{021011}
	(\bibinfo{year}{2020}).
	
	\bibitem{Zhang2021}
	\bibinfo{author}{Zhang, L.} \emph{et~al.}
	\newblock \bibinfo{title}{Van der {{Waals}} heterostructure polaritons with
		moir{\'e}-induced nonlinearity}.
	\newblock \emph{\bibinfo{journal}{Nature}} \textbf{\bibinfo{volume}{591}},
	\bibinfo{pages}{61--65} (\bibinfo{year}{2021}).
	
	\bibitem{Scherzer2024}
	\bibinfo{author}{Scherzer, J.} \emph{et~al.}
	\newblock \bibinfo{title}{Correlated magnetism of moir{\'e} exciton-polaritons
		on a triangular electron-spin lattice}.
	\newblock \emph{\bibinfo{journal}{arXiv.org,}} \bibinfo{pages}{2405.12698}
	(\bibinfo{year}{2024}).
	
	\bibitem{Schneider2016}
	\bibinfo{author}{Schneider, C.} \emph{et~al.}
	\newblock \bibinfo{title}{Exciton-polariton trapping and potential landscape
		engineering}.
	\newblock \emph{\bibinfo{journal}{Reports on Progress in Physics}}
	\textbf{\bibinfo{volume}{80}}, \bibinfo{pages}{016503}
	(\bibinfo{year}{2016}).
	
	\bibitem{Klembt2018}
	\bibinfo{author}{Klembt, S.} \emph{et~al.}
	\newblock \bibinfo{title}{Exciton-polariton topological insulator}.
	\newblock \emph{\bibinfo{journal}{Nature}} \textbf{\bibinfo{volume}{562}},
	\bibinfo{pages}{552--556} (\bibinfo{year}{2018}).
	
	\bibitem{Fontaine2022}
	\bibinfo{author}{Fontaine, Q.} \emph{et~al.}
	\newblock \bibinfo{title}{Kardar-{{Parisi-Zhang}} universality in a
		one-dimensional polariton condensate}.
	\newblock \emph{\bibinfo{journal}{Nature}} \textbf{\bibinfo{volume}{608}},
	\bibinfo{pages}{687--691} (\bibinfo{year}{2022}).
	
	\bibitem{Borghardt2017}
	\bibinfo{author}{Borghardt, S.} \emph{et~al.}
	\newblock \bibinfo{title}{Engineering of optical and electronic band gaps in
		transition metal dichalcogenide monolayers through external dielectric
		screening}.
	\newblock \emph{\bibinfo{journal}{Physical Review Materials}}
	\textbf{\bibinfo{volume}{1}}, \bibinfo{pages}{054001} (\bibinfo{year}{2017}).
	
	\bibitem{Raja2017}
	\bibinfo{author}{Raja, A.} \emph{et~al.}
	\newblock \bibinfo{title}{Coulomb engineering of the bandgap and excitons in
		two-dimensional materials}.
	\newblock \emph{\bibinfo{journal}{Nature Communications}}
	\textbf{\bibinfo{volume}{8}}, \bibinfo{pages}{15251} (\bibinfo{year}{2017}).
	
	\bibitem{Peimyoo2020}
	\bibinfo{author}{Peimyoo, N.} \emph{et~al.}
	\newblock \bibinfo{title}{Engineering {{Dielectric Screening}} for
		{{Potential-well Arrays}} of {{Excitons}} in {{2D Materials}}}.
	\newblock \emph{\bibinfo{journal}{ACS Applied Materials \& Interfaces}}
	\textbf{\bibinfo{volume}{12}}, \bibinfo{pages}{55134--55140}
	(\bibinfo{year}{2020}).
	
	\bibitem{BenMhenni2025}
	\bibinfo{author}{Ben~Mhenni, A.} \emph{et~al.}
	\newblock \bibinfo{title}{Breakdown of the {{Static Dielectric Screening
				Approximation}} of {{Coulomb Interactions}} in {{Atomically Thin
				Semiconductors}}}.
	\newblock \emph{\bibinfo{journal}{ACS Nano}} \textbf{\bibinfo{volume}{19}},
	\bibinfo{pages}{4269--4278} (\bibinfo{year}{2025}).
	
	\bibitem{Wurdack2021}
	\bibinfo{author}{Wurdack, M.} \emph{et~al.}
	\newblock \bibinfo{title}{Motional narrowing, ballistic transport, and trapping
		of room-temperature exciton polaritons in an atomically-thin semiconductor}.
	\newblock \emph{\bibinfo{journal}{Nature Communications}}
	\textbf{\bibinfo{volume}{12}}, \bibinfo{pages}{5366} (\bibinfo{year}{2021}).
	
	\bibitem{Rupprecht2020}
	\bibinfo{author}{Rupprecht, C.} \emph{et~al.}
	\newblock \bibinfo{title}{Demonstration of a polariton step potential by local
		variation of light-matter coupling in a van-der-{{Waals}} heterostructure}.
	\newblock \emph{\bibinfo{journal}{Optics Express}}
	\textbf{\bibinfo{volume}{28}}, \bibinfo{pages}{18649--18657}
	(\bibinfo{year}{2020}).
	
	\bibitem{Wurdack2022}
	\bibinfo{author}{Wurdack, M.} \emph{et~al.}
	\newblock \bibinfo{title}{Enhancing {{Ground-State Population}} and
		{{Macroscopic Coherence}} of {{Room-Temperature WS}}{\textsubscript{2}}
		{{Polaritons}} through {{Engineered Confinement}}}.
	\newblock \emph{\bibinfo{journal}{Physical Review Letters}}
	\textbf{\bibinfo{volume}{129}}, \bibinfo{pages}{147402}
	(\bibinfo{year}{2022}).
	
	\bibitem{Li2023}
	\bibinfo{author}{Li, D.} \emph{et~al.}
	\newblock \bibinfo{title}{Trapping-induced quantum beats in a van-der-{{Waals}}
		heterostructure microcavity observed by two-dimensional micro-spectroscopy}.
	\newblock \emph{\bibinfo{journal}{Optical Materials Express}}
	\textbf{\bibinfo{volume}{13}}, \bibinfo{pages}{2798--2807}
	(\bibinfo{year}{2023}).
	
	\bibitem{Zhao2021}
	\bibinfo{author}{Zhao, J.} \emph{et~al.}
	\newblock \bibinfo{title}{Ultralow {{Threshold Polariton Condensate}} in a
		{{Monolayer Semiconductor Microcavity}} at {{Room Temperature}}}.
	\newblock \emph{\bibinfo{journal}{Nano Letters}} \textbf{\bibinfo{volume}{21}},
	\bibinfo{pages}{3331--3339} (\bibinfo{year}{2021}).
	
	\bibitem{Blais2004}
	\bibinfo{author}{Blais, A.}, \bibinfo{author}{Huang, R.-S.},
	\bibinfo{author}{Wallraff, A.}, \bibinfo{author}{Girvin, S.~M.} \&
	\bibinfo{author}{Schoelkopf, R.~J.}
	\newblock \bibinfo{title}{Cavity quantum electrodynamics for superconducting
		electrical circuits: {{An}} architecture for quantum computation}.
	\newblock \emph{\bibinfo{journal}{Physical Review A}}
	\textbf{\bibinfo{volume}{69}}, \bibinfo{pages}{062320}
	(\bibinfo{year}{2004}).
	
	\bibitem{Majer2007}
	\bibinfo{author}{Majer, J.} \emph{et~al.}
	\newblock \bibinfo{title}{Coupling superconducting qubits via a cavity bus}.
	\newblock \emph{\bibinfo{journal}{Nature}} \textbf{\bibinfo{volume}{449}},
	\bibinfo{pages}{443--447} (\bibinfo{year}{2007}).
	
	\bibitem{Ritsch2013}
	\bibinfo{author}{Ritsch, H.}, \bibinfo{author}{Domokos, P.},
	\bibinfo{author}{Brennecke, F.} \& \bibinfo{author}{Esslinger, T.}
	\newblock \bibinfo{title}{Cold atoms in cavity-generated dynamical optical
		potentials}.
	\newblock \emph{\bibinfo{journal}{Reviews of Modern Physics}}
	\textbf{\bibinfo{volume}{85}}, \bibinfo{pages}{553--601}
	(\bibinfo{year}{2013}).
	
	\bibitem{Evans2018}
	\bibinfo{author}{Evans, R.~E.} \emph{et~al.}
	\newblock \bibinfo{title}{Photon-mediated interactions between quantum emitters
		in a diamond nanocavity}.
	\newblock \emph{\bibinfo{journal}{Science}} \textbf{\bibinfo{volume}{362}},
	\bibinfo{pages}{662--665} (\bibinfo{year}{2018}).
	
	\bibitem{Vijayan2024}
	\bibinfo{author}{Vijayan, J.} \emph{et~al.}
	\newblock \bibinfo{title}{Cavity-mediated long-range interactions in levitated
		optomechanics}.
	\newblock \emph{\bibinfo{journal}{Nature Physics}}
	\textbf{\bibinfo{volume}{20}}, \bibinfo{pages}{859--864}
	(\bibinfo{year}{2024}).
	
	\bibitem{Vadia2021}
	\bibinfo{author}{Vadia, S.} \emph{et~al.}
	\newblock \bibinfo{title}{Open-{{Cavity}} in {{Closed-Cycle Cryostat}} as a
		{{Quantum Optics Platform}}}.
	\newblock \emph{\bibinfo{journal}{PRX Quantum}} \textbf{\bibinfo{volume}{2}},
	\bibinfo{pages}{040318} (\bibinfo{year}{2021}).
	
	\bibitem{Tan2023}
	\bibinfo{author}{Tan, L.~B.} \emph{et~al.}
	\newblock \bibinfo{title}{Bose {{Polaron Interactions}} in a {{Cavity-Coupled
				Monolayer Semiconductor}}}.
	\newblock \emph{\bibinfo{journal}{Physical Review X}}
	\textbf{\bibinfo{volume}{13}}, \bibinfo{pages}{031036}
	(\bibinfo{year}{2023}).
	
	\bibitem{Savona1995}
	\bibinfo{author}{Savona, V.}, \bibinfo{author}{Andreani, L.},
	\bibinfo{author}{Schwendimann, P.} \& \bibinfo{author}{Quattropani, A.}
	\newblock \bibinfo{title}{Quantum well excitons in semiconductor microcavities:
		{{Unified}} treatment of weak and strong coupling regimes}.
	\newblock \emph{\bibinfo{journal}{Solid State Communications}}
	\textbf{\bibinfo{volume}{93}}, \bibinfo{pages}{733--739}
	(\bibinfo{year}{1995}).
	
	\bibitem{Rupprecht2021}
	\bibinfo{author}{Rupprecht, C.} \emph{et~al.}
	\newblock \bibinfo{title}{Micro-mechanical assembly and characterization of
		high-quality {{Fabry-P{\'e}rot}} microcavities for the integration of
		two-dimensional materials}.
	\newblock \emph{\bibinfo{journal}{Applied Physics Letters}}
	\textbf{\bibinfo{volume}{118}} (\bibinfo{year}{2021}).
	
	\bibitem{Thureja2022}
	\bibinfo{author}{Thureja, D.} \emph{et~al.}
	\newblock \bibinfo{title}{Electrically tunable quantum confinement of neutral
		excitons}.
	\newblock \emph{\bibinfo{journal}{Nature}} \textbf{\bibinfo{volume}{606}},
	\bibinfo{pages}{298--304} (\bibinfo{year}{2022}).
	
	\bibitem{Hu2024}
	\bibinfo{author}{Hu, J.} \emph{et~al.}
	\newblock \bibinfo{title}{Quantum control of exciton wave functions in {{2D}}
		semiconductors}.
	\newblock \emph{\bibinfo{journal}{Science Advances}}
	\textbf{\bibinfo{volume}{10}}, \bibinfo{pages}{eadk6369}
	(\bibinfo{year}{2024}).
	
	\bibitem{Gu2021}
	\bibinfo{author}{Gu, J.} \emph{et~al.}
	\newblock \bibinfo{title}{Enhanced nonlinear interaction of polaritons via
		excitonic {{Rydberg}} states in monolayer {{WSe}}{\textsubscript{2}}}.
	\newblock \emph{\bibinfo{journal}{Nature Communications}}
	\textbf{\bibinfo{volume}{12}}, \bibinfo{pages}{2269} (\bibinfo{year}{2021}).
	
	\bibitem{Gu2024}
	\bibinfo{author}{Gu, L.} \emph{et~al.}
	\newblock \bibinfo{title}{Giant optical nonlinearity of {{Fermi}} polarons in
		atomically thin semiconductors}.
	\newblock \emph{\bibinfo{journal}{Nature Photonics}}
	\textbf{\bibinfo{volume}{18}}, \bibinfo{pages}{816--822}
	(\bibinfo{year}{2024}).
	
	\bibitem{Polovnikov2024}
	\bibinfo{author}{Polovnikov, B.} \emph{et~al.}
	\newblock \bibinfo{title}{Field-{{Induced Hybridization}} of {{Moir{\'e}
				Excitons}} in {{MoSe}}{\textsubscript{2}}/{{WS}}{\textsubscript{2}}
		{{Heterobilayers}}}.
	\newblock \emph{\bibinfo{journal}{Physical Review Letters}}
	\textbf{\bibinfo{volume}{132}}, \bibinfo{pages}{076902}
	(\bibinfo{year}{2024}).
	
	\bibitem{Boulier2020}
	\bibinfo{author}{Boulier, T.} \emph{et~al.}
	\newblock \bibinfo{title}{Microcavity {{Polaritons}} for {{Quantum
				Simulation}}}.
	\newblock \emph{\bibinfo{journal}{Advanced Quantum Technologies}}
	\textbf{\bibinfo{volume}{3}}, \bibinfo{pages}{2000052}
	(\bibinfo{year}{2020}).
	
	\bibitem{Son2020}
	\bibinfo{author}{Son, S.} \emph{et~al.}
	\newblock \bibinfo{title}{Strongly adhesive dry transfer technique for van der
		{{Waals}} heterostructure}.
	\newblock \emph{\bibinfo{journal}{2D Materials}} \textbf{\bibinfo{volume}{7}},
	\bibinfo{pages}{041005} (\bibinfo{year}{2020}).
	
	\bibitem{Shin2021}
	\bibinfo{author}{Shin, M.~J.} \& \bibinfo{author}{Shin, Y.~J.}
	\newblock \bibinfo{title}{Stacking of {{2D}} materials containing a thin layer
		of hexagonal boron nitride using polycaprolactone}.
	\newblock \emph{\bibinfo{journal}{Journal of the Korean Physical Society}}
	\textbf{\bibinfo{volume}{78}}, \bibinfo{pages}{1089--1094}
	(\bibinfo{year}{2021}).
	
	\bibitem{Tabataba-Vakili2024}
	\bibinfo{author}{{Tabataba-Vakili}, F.} \emph{et~al.}
	\newblock \bibinfo{title}{Metasurface of {{Strongly Coupled Excitons}} and
		{{Nanoplasmonic Arrays}}}.
	\newblock \emph{\bibinfo{journal}{Nano Letters}} \textbf{\bibinfo{volume}{24}},
	\bibinfo{pages}{10090--10097} (\bibinfo{year}{2024}).
	
	\bibitem{Mader2022}
	\bibinfo{author}{Mader, M.}, \bibinfo{author}{Benedikter, J.},
	\bibinfo{author}{Husel, L.}, \bibinfo{author}{H{\"a}nsch, T.~W.} \&
	\bibinfo{author}{Hunger, D.}
	\newblock \bibinfo{title}{Quantitative determination of the complex
		polarizability of individual nanoparticles by scanning cavity microscopy}.
	\newblock \emph{\bibinfo{journal}{ACS Photonics}} \textbf{\bibinfo{volume}{9}},
	\bibinfo{pages}{466--473} (\bibinfo{year}{2022}).
\end{thebibliography}
\end{document}


\title{Supplementary Information: Cavity-mediated exciton hopping in a dielectrically engineered polariton system}

\author{Lukas Husel}
\def\LMU{Fakult\"at f\"ur Physik, Munich Quantum Center, and Center for NanoScience (CeNS), Ludwig-Maximilians-Universit\"at M\"unchen, Geschwister-Scholl-Platz~1, D-80539 M\"unchen, Germany}
\affiliation{\LMU}

\author{Farsane Tabataba-Vakili}
\affiliation{\LMU}
\affiliation{Munich Center for Quantum Science and Technology (MCQST), Schellingstr.~4, D-80799 M\"unchen, Germany} 
\affiliation{Institute of Condensed Matter Physics, Technische Universität Braunschweig, 38106 Braunschweig, Germany }

\author{Johannes Scherzer}
\affiliation{\LMU}

\author{Lukas Krelle}
\affiliation{\LMU}
\def\DAR{Present affiliation: Institute for Condensed Matter Physics, TU Darmstadt, Hochschulstr.~6-8, D-64289 Darmstadt, Germany}
\affiliation{\DAR}

\author{Ismail Bilgin}
\affiliation{\LMU}

\author{Samarth Vadia}
\affiliation{\LMU}

\author{Kenji Watanabe}
\affiliation{Research Center for Electronic and Optical Materials, National Institute for Materials Science, 1-1 Namiki, Tsukuba 305-0044, Japan}

\author{Takashi Taniguchi}
\affiliation{Research Center for Materials Nanoarchitectonics, National Institute for Materials Science, 1-1 Namiki, Tsukuba 305-0044, Japan}

\author{Iacopo Carusotto}
\affiliation{Pitaevskii BEC Center, INO-CNR and Dipartimento di Fisica, Universita di Trento, via Sommarive 14, I-38123 Trento, Italy}  

\author{Alexander H\"ogele}
\affiliation{\LMU}
\affiliation{Munich Center for Quantum Science and Technology (MCQST), Schellingstr.~4, D-80799 M\"unchen, Germany}

\date{\today}

\maketitle

\section*{Supplementary Note I: Confocal photoluminescence spectroscopy}
\label{sec:SupSec1}

To investigate the effect of the engineered dielectric environment on the exciton resonance energies, we performed confocal cryogenic photoluminescence (PL) spectroscopy on the fabricated device. For this measurement, the planar cavity mirror with the van der Waals heterostack on top was mounted in backscattering geometry inside a closed-cycle cryostat held at 4.3~K. The CW pump laser had a wavelength of 675~nm, the spatial resolution in the transverse direction was approx. 1~$\mu$m. 

A raster-scan map of integrated PL for the representative etch site area P$_2$ is shown in Supplementary Fig.~\ref{figS1}a. In Supplementary Fig.~\ref{figS1}b, we plot a spectrally resolved PL linecut across the etch site pair, measured along the dashed line in Supplementary Fig.~\ref{figS1}a. The PL spectra are dominated by the resonance of excitons in the encapsulated monolayer region, and exhibit a redshifted shoulder at the position of the etch sites, with energetic positions indicated by white dots. This additional resonance originates from excitons localized to the air-hole area, with exciton binding energy and TMD band gap modified by the local change in dielectric environment~\cite{Khestanova2024, Raja2017, Peimyoo2020, Borghardt2017, BenMhenni2025}. The resonance originating from the fully encapsulated domain, at maximum PL intensity in Supplementary Fig.~\ref{figS1}b, exhibits a pronounced blueshift at the etch site position, which likely originates from local strain~\cite{Schmidt2016} induced during fabrication. This blueshift is irrelevant to any claims made in the main text, as well as for the computation of lower polariton energies in Supplementary Note~IV and cavity-mediated exciton-hopping in Supplementary Note~V.

\begin{figure}
	\centering
	\includegraphics[scale=1]{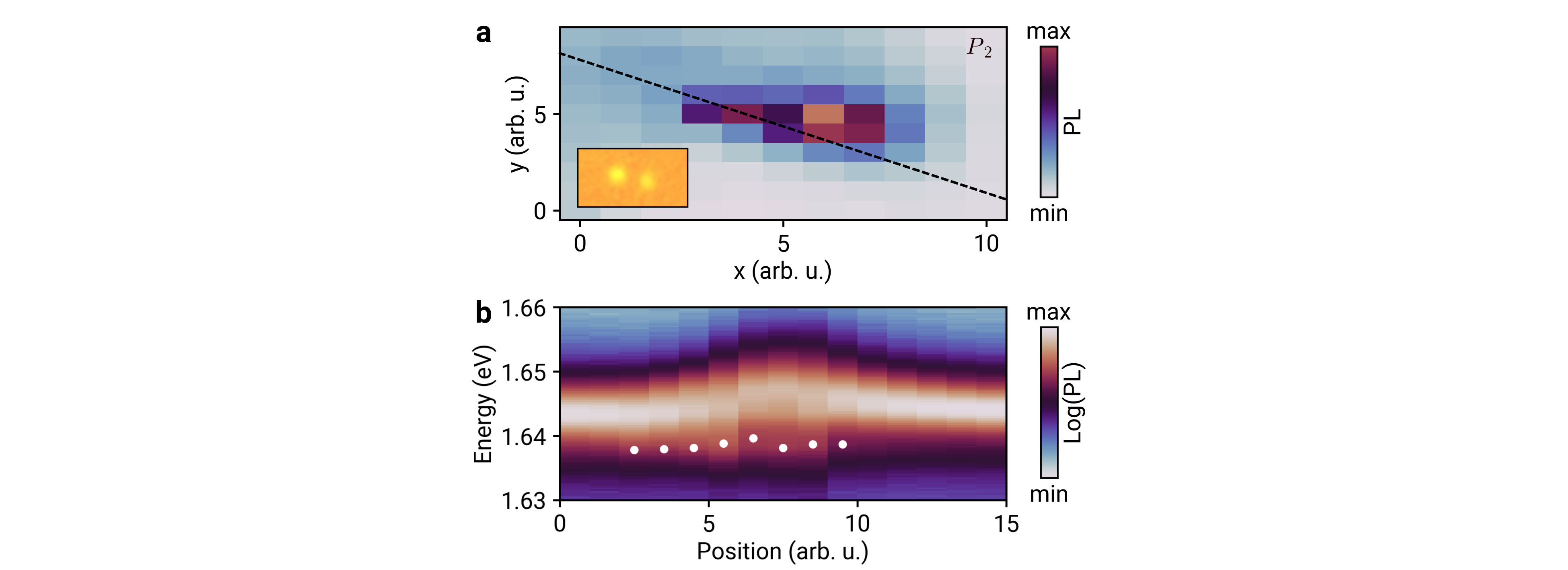}
	\caption{\textbf{Confocal photoluminescence spectroscopy.} \textbf{a}, Raster-scan map of etch site pair P$_2$ in cryogenic confocal PL, integrated in the spectral range 1.61 to 1.72~eV. The inset shows a microscope image of the investigated device area. \textbf{b}, Spectrally resolved PL-linecut along the dashed line in~\textbf{a}. The white dots are energetic positions of a redshifted shoulder, obtained from the fit of double Gaussian profiles to the PL spectra at each position. }
	\label{figS1}
\end{figure}

\section*{Supplementary Note II: Analysis of cavity transmission spectra}
\label{sec:SupSec2}
\begin{figure}[t]
	\centering
	\includegraphics[scale=1]{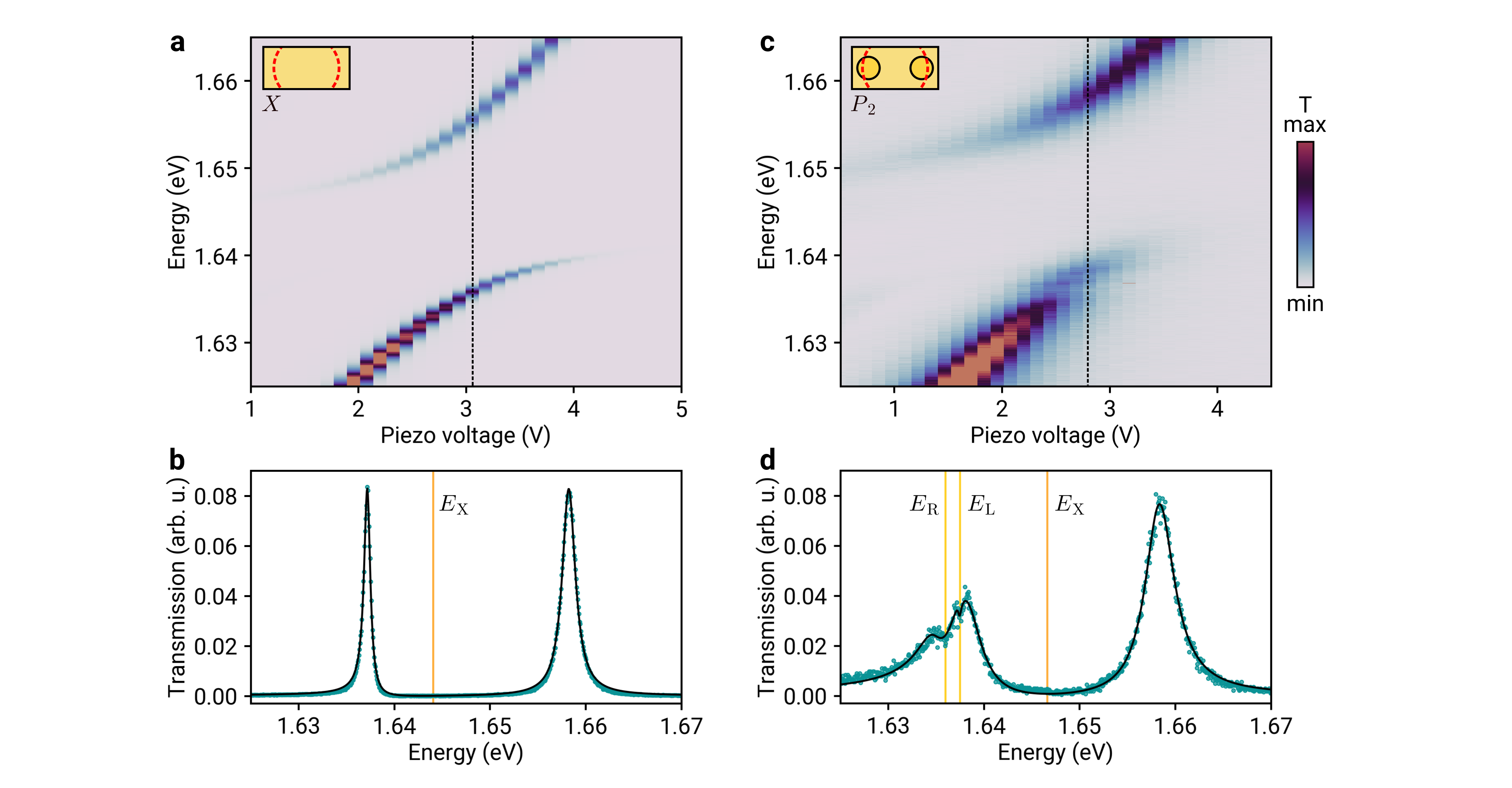}
	\caption{\textbf{Analysis of cavity transmission spectra.} \textbf{a}, Cavity transmission as function of cavity length, tuned via piezoelectric actuator voltage. The cavity mode is positioned on a TMD area fully encapsulated by hBN. \textbf{b}, Experimental transmission spectrum obtained for the voltage marked by the vertical dashed line in \textbf{a} (green dots), along with a fit of the model of Eq.~\ref{Eq:T} for a single cavity-coupled exciton resonance at energy $E_\mathrm{X}$ (solid line). \textbf{c}, Cavity length sweep with the cavity positioned at the etch site pair P$_2$, resulting in additional cavity-coupled excitonic resonances originating from the etch site domains. \textbf{d}, Same as \textbf{b}, but with a model fit for three cavity-coupled exciton resonances of different energies and coupling strengths. Resonance energies of excitons localized to the etch site domains are labeled $E_\mathrm{L}$ and $E_\mathrm{R}$, $E_\mathrm{X}$ is the energy of excitons in the hBN encapsulated device area. All data in the figure are normalized to the maximum transmission measured in the respective cavity length sweep.}
	\label{figS2}
\end{figure}
To determine exciton resonance energies and light-matter coupling strengths from cavity-based measurements, we evoke a dissipative model for the cavity transmission. The starting point is the time-independent Hamiltonian of the system, which describes excitons of different energies $E_{i}$ localized to non-overlapping areas of the device labeled $i$ as individual quantum wells coupled to a single cavity mode, each with a light-matter coupling strength $g_i$,
\begin{equation}
	H = E_C a^{\dagger}a + \sum_i E_{i} b_i^{\dagger}b_i + \hbar g_i(b_i^{\dagger}a+a^{\dagger}b_i).
	\label{Eq:H}
\end{equation}
In this expression, $b_i$ is the bosonic annihilation operator of excitons in area $i$, while $a$ is the bosonic annihilation operator for cavity photons at energy $E_C$. Combining Eq.~\ref{Eq:H} with the dissipative input-output formalism of Ref.~\cite{Evans2018}, we find the cavity transmission at energy $E$
\begin{equation}
	T(E) = \eta \kappa_{\mathrm{m}}^2 \left| \mathrm{i}(E_C-E) + \hbar\kappa/2 + \sum_i \frac{(\hbar g_i)^2}{\mathrm{i}(E_{i}-E)+\hbar\Gamma_i/2} \right|^{-2}, 
	\label{Eq:T}
\end{equation}
in which $\kappa$ and $\Gamma_i$ are the cavity and exciton FWHM linewidths, respectively, and $\eta$ is a proportionality constant. $\kappa_{\mathrm{m}} = \kappa_{\mathrm{m, in}} \kappa_{\mathrm{m, out}}$ is the product of decay rates through the in- and outcoupling mirror, $\kappa_{\mathrm{m, in}}$ and $\kappa_{\mathrm{m, out}}$, respectively. As explained in the Methods section, the linewidth of our cavity is broadened by vibrational fluctuations. By contrast, the light-matter coupling dynamics are dominated by the Lorentzian contribution to the cavity linewidth as the frequency of the mechanical vibrations is orders of magnitude slower than the energy exchange rate between the cavity and excitons. In practice, we found this distinction to have negligible impact on the data analysis procedure presented in the following, such that vibration broadening is implicitly accounted for in the value for $\kappa$ in Eq.~\ref{Eq:T}.

In the experiment, we measure cavity transmission as a function of cavity length, with a typical result shown in Supplementary Fig.~\ref{figS2}a for the cavity mode positioned on the fully encapsulated TMD area. All spectra are background-corrected for a constant CCD offset and normalized with respect to the maximum measured transmission for a given length sweep. From transmission measurements of the empty cavity (covering only two hBN layers), we found that $\kappa_{\mathrm{m}}$ varied by about 20\% in the investigated spectral range, a result of wavelength-dependent mirror reflectivity. To account for this effect, we normalized the measured transmission spectra $T_{\mathrm{meas}}(E)$ by the values $\kappa_{\mathrm{m, meas}}^2 (E)$ measured for the empty cavity, which we found to improve the results of the fit procedure described in the following.

To determine $E_{i}$ and $g_i$, we fit Eq.~\ref{Eq:T} to normalized transmission spectra at different cavity lengths. Data and a representative fit result are shown in Supplementary Fig.~\ref{figS2}b by green dots and the black line, respectively. The fit yields good agreement between model and data, also in the case of multiple exciton domains coupled to the cavity, for which representative data and fit results are shown in Supplementary Figs.~\ref{figS2}c and d, respectively. 

For each excitonic resonance $i$ (identified by the presence of corresponding polariton branches), we compute mean values $\bar{E_i}$, $\bar{g_i}$ and standard error $\delta E_i$, $\delta g_i$ from fit results for $E_i$ and $g_i$ obtained at different cavity lengths. $\bar{E_i} \pm \delta E_i$ and $\bar{g_i} \pm \delta g_i$ constitute our measurement results and uncertainty. To obtain good fit quality, we ensured that the fit results obtained for individual cavity lengths complied with three criteria: First, the fit results for the cavity energy should be a linear function of cavity length. Fits for which the obtained cavity energy deviated strongly from this linear dependence were discarded in the analysis. Second, only fits in which the cavity energy was in or near resonance with the exciton energy were used, a condition which we found to minimize the errors in the fit parameters of interest. Third, fits whose cost functions deviated largely from those obtained for similar cavity lengths were discarded. Fit results for different cavity lengths were used to compute $\bar{E_i} \pm \delta E_i$ and $\bar{g_i} \pm \delta g_i$, with the precise number of spectra available for analysis determined by the signal to noise ratio in the respective polariton branches.

\begin{figure}[t]
	\centering
	\includegraphics[scale=1]{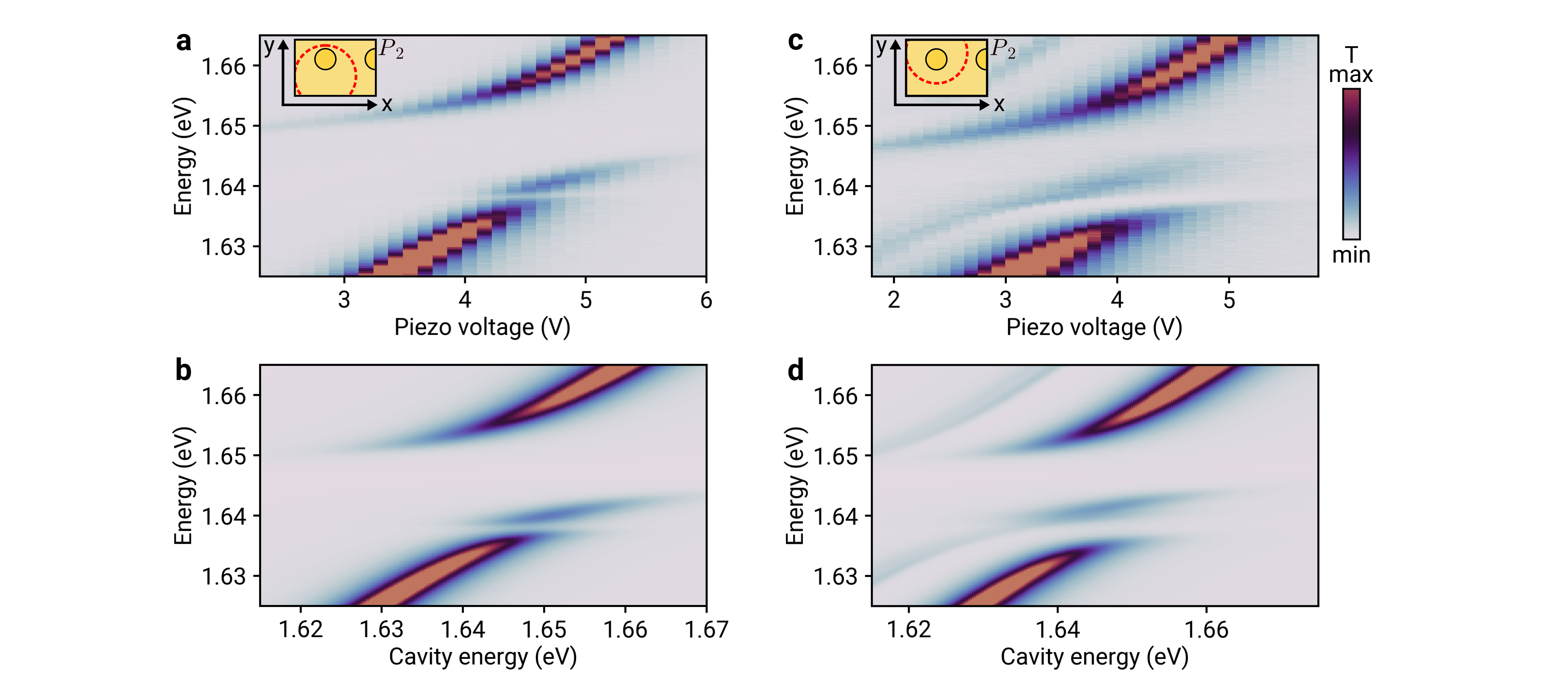}
	\caption{\textbf{Higher order transverse cavity modes.} \textbf{a}, Measured cavity transmission for a sweep of the cavity length. The position of the cavity mode with respect to the etch site pair P$_2$ is illustrated in the inset. Excitons in the left etch site and the encapsulated monolayer couple to the cavity, resulting in three polariton branches. \textbf{b}, Plot of theoretical cavity transmission (Eq.~\ref{Eq:T}), with parameters obtained from a fit to the data in in \textbf{a}. \textbf{c}, Same as in \textbf{a}, with a different cavity mode position as illustrated in the inset. Higher order transverse cavity modes contribute additional cavity resonances to the spectra. \textbf{d}, Plot of theoretical cavity transmission, with the model of Eq.~\ref{Eq:T} extended to the case of three cavity modes coupling to the excitonic resonances. The model parameters were adjusted to yield agreement with the data in \textbf{c}.}
	\label{figS3}
\end{figure}

Due to an ellipticity in the fiber mirror profile, the cavity exhibits non-degenerate higher order Hermite-Gaussian modes~\cite{Benedikter2015}, which contribute to the measured transmission spectra. We found this contribution to depend on the position of the cavity mode relative to the etch sites, a finding we illustrate in Supplementary Fig.~\ref{figS3}. If the center of the cavity mode was placed near the bottom of the etch site center (coordinate system defined in the inset of Supplementary Fig.~\ref{figS3}a), the transmission spectra were typically dominated by a single cavity resonance associated with the fundamental Gaussian cavity mode, as illustrated by the data shown in Supplementary Figs.~\ref{figS3}a. Results of the theoretical model of Eq.~\ref{Eq:T}, plotted in Supplementary Figs.~\ref{figS3}b, yield excellent agreement with the data.  

If the cavity mode was moved close to the etch site center, with representative data shown in Supplementary Fig.~\ref{figS3}c, higher order modes contributed cavity resonances to the measured spectra, blueshifted with respect to the fundamental mode. To support this assignment, we extend the model of Eq.~\ref{Eq:T} to describe exciton-coupling to three different cavity modes. The result is shown in Supplementary Fig.~\ref{figS3}d and agrees well with the measurement result.  

The contribution of the higher order modes to the measured spectra persisted as the cavity mode was moved towards the top of the etch site center. This asymmetry, which was also observed for the etch site pairs P$_1$ and P$_3$, is likely a result of asymmetric transversal mode profiles caused by a tilted cavity fiber. For several positions close to the etch site centers, the higher order modes were resonant with the middle polariton branches, adding uncertainty to the fit procedure described above. To determine exciton coupling strengths and resonance energies, as well as to visualize cavity-mediated exciton hopping in Figs.~3 and 4 of the main text, we therefore restricted our measurements to cavity mode positions with negligible higher order mode contributions, as indicated in the insets of the respective figure panels. In the measurements of local polariton energy shifts as shown in Fig.~2 of the main text, higher order modes brightened the middle polariton branches, an effect which is negligible for the analysis in Supplementary Note~IV.

\section*{Supplementary Note III: Cavity-coupled exciton domains}
\label{sec:SupSec3}
\begin{figure}[t]
	\centering
	\includegraphics[scale=1]{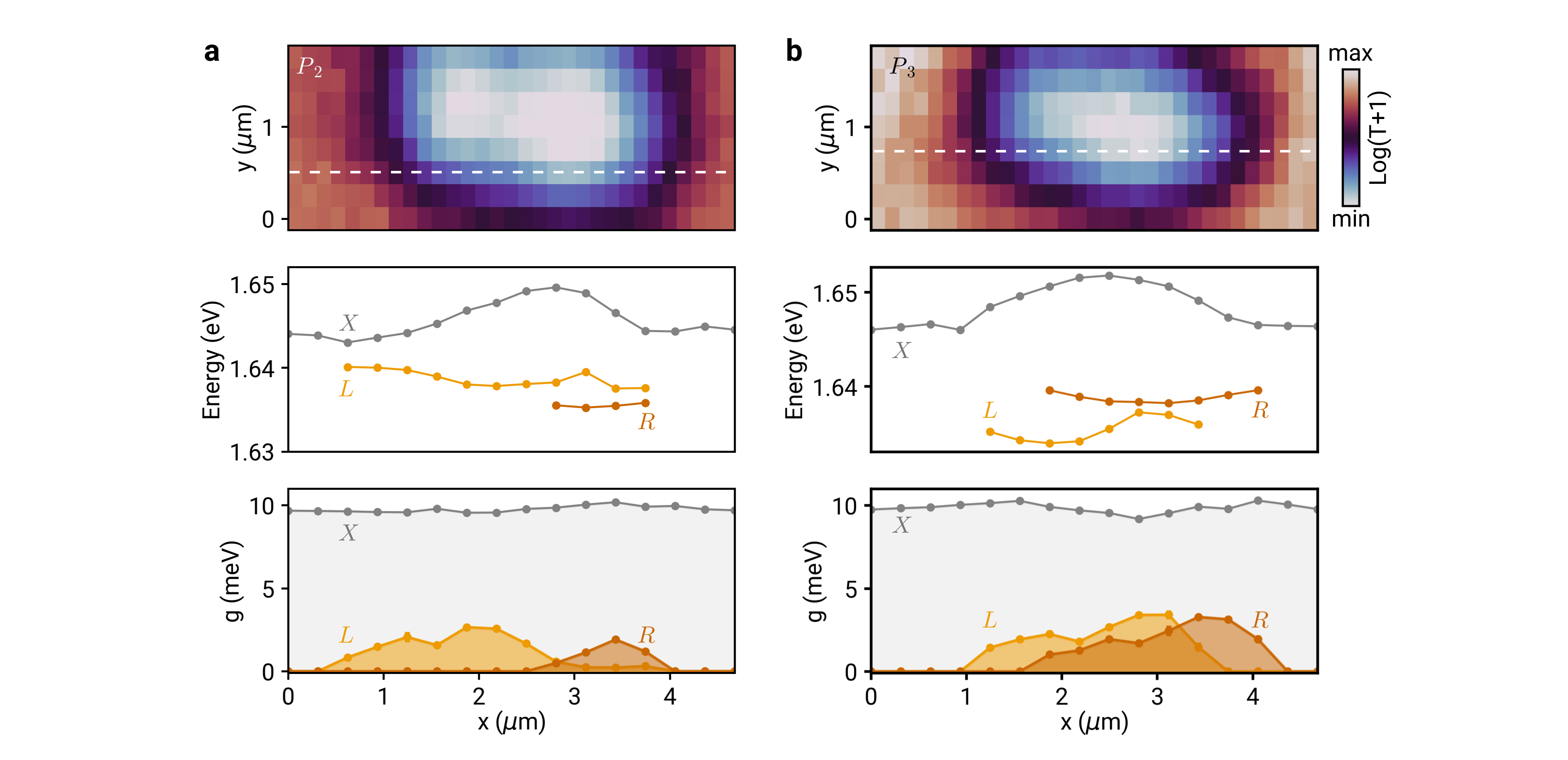}
	\caption{\textbf{Cavity-coupled exciton domains.} \textbf{a}, Top panel: Cavity transmission map at fixed cavity energy ($E_C$ = 1.612~eV) of the etch site pair P$_2$. Axes of the coordinate system are the same as in Fig.~1b of the main text. Middle panel: Exciton resonance energies along the white dashed line in the top panel, identified from cavity length sweeps and originating from $X$, $L$ and $R$ domains as defined in the main text. Bottom panel: Exciton light matter coupling strengths $g$ for the resonances identified in the middle panel. At positions where polariton branches associated with the individual resonances were absent in the transmission spectra, no data points are shown for the energies, and the values of $g$ are set to zero. \textbf{b}, Same as \textbf{a} but for the etch site pair P$_3$. The cavity transmission map was obtained at $E_C$ = 1.666~eV.} 
	\label{figS4}
\end{figure}

To investigate the properties of dielectrically engineered exciton domains, we harnessed the tunability of our open cavity system. The top panel in Supplementary Fig.~\ref{figS4}a shows a raster-scan map of cavity transmission for the etch site pair P$_2$, obtained at a fixed cavity energy spectrally detuned from any exciton resonances. At this energy, the etch site induces mainly photonic scattering loss to the cavity mode, resulting in the observed decrease in cavity transmission.

We performed cavity length sweeps at each position along the dashed line in the transmission map of Supplementary Fig.~\ref{figS4}a. Using the analysis procedure described in Supplementary Note~II, we determined resonance energy $E_i$ and light-matter coupling strength $g_i$ of all exciton resonances $i$ which contributed polariton branches to the transmission spectra. The results for $E_i$ and $g_i$ are shown in the middle and bottom panels of Supplementary Fig.~\ref{figS4}a, respectively. Results for $E$ and $g$ at each position were labeled according to the corresponding exciton resonances $X$, $L$ and $R$ as shown in the figure.

As the cavity is placed at the left side of the etch site ($x=0 \ \mu$m, left side in Supplementary Fig.~\ref{figS4}a), we find a single exciton resonance labeled $X$ with light-matter coupling strength $g_{X}$ = 9.7~meV, which stems from the TMD area fully encapsulated with hBN. Its resonance energy experiences a blueshift as the cavity mode is moved towards the etch site centers, which is consistent with data obtained in confocal PL and likely a result of strain, as discussed in Supplementary Note~I. 

As expected, we find two exciton domains defined by the through holes, giving rise to resonances $L$ and $R$ with maximum values of light-matter coupling strength at the respective etch site centers. The maximum coupling strength $g_{L/R}$ for these domains is expected to scale as $g_{L/R} \propto \sqrt{\eta_A}$~\cite{Thureja2023, Gebhardt2019} with $\eta_A$ the overlap between exciton domain and cavity mode. Using the domain area $A_S$ and the 1/$\mathrm{e}^2$ area of the transverse cavity field $A_C$, we estimate $\mathrm{max}(g_{XL})\sqrt{A_S/A_C}$ = 2.91~meV, close to the maximum measured value $g_{L} = 2.65 \pm 0.04$~meV in Supplementary Fig.~\ref{figS4}b. The difference in energy and coupling strengths between the $L$ and $R$ domains reflect typical inhomogeneities in TMD-based van der Waals heterostructures. On the length scale of the cavity mode waist, the spatial variations in exciton energy are on the order of the typical exciton linewidth for all three resonances in Supplementary Fig.~4. As a result, these variations will add to the inhomogeneous broadening of the exciton linewidth. In our analysis, we therefore treat each exciton in Supplementary Fig.~4 as an inhomogeneously broadened resonance with a single, spatially dependent frequency.

Repeating measurements and analysis for the etch site pair P$_3$, with data shown in Supplementary Fig.~\ref{figS4}c, yields similar results as for P$_2$. Again, we find different excitonic domains $L$ and $R$ defined by the air holes, whose coupling strength maxima are observed at smaller distance than for P$_2$ due to the reduced etch site distance. 

Our observation of redshifted exciton resonaces at reduced dielectric screening is in agreement with recent results~\cite{BenMhenni2025, Khestanova2024}. Notably, it contrasts the case of graphene-encapsulated TMD monolayers~\cite{Raja2017}, for which the exciton energy is known to blueshift at reduced dielectric screening, which is a result of the difference in dielectric response of graphene and hBN~\cite{BenMhenni2025}. For the sake of completeness, we note that we observed spectrally broad, blueshifted exciton resonances of unknown origin for a small number of etch sites, which coupled weakly to the cavity and are irrelevant to the measurements presented in the following. We summarize that our fabrication method allows to deterministically create exction domains redshifted in resonance energy by up to 10~meV from excitons in the hBN encapsulated monolayer area, with coupling strengths of approx. 2.5~meV defined by the domain size.

\section*{Supplementary Note IV: Local polariton energy shifts}

\begin{figure}[t]
	\centering
	\includegraphics[scale=1]{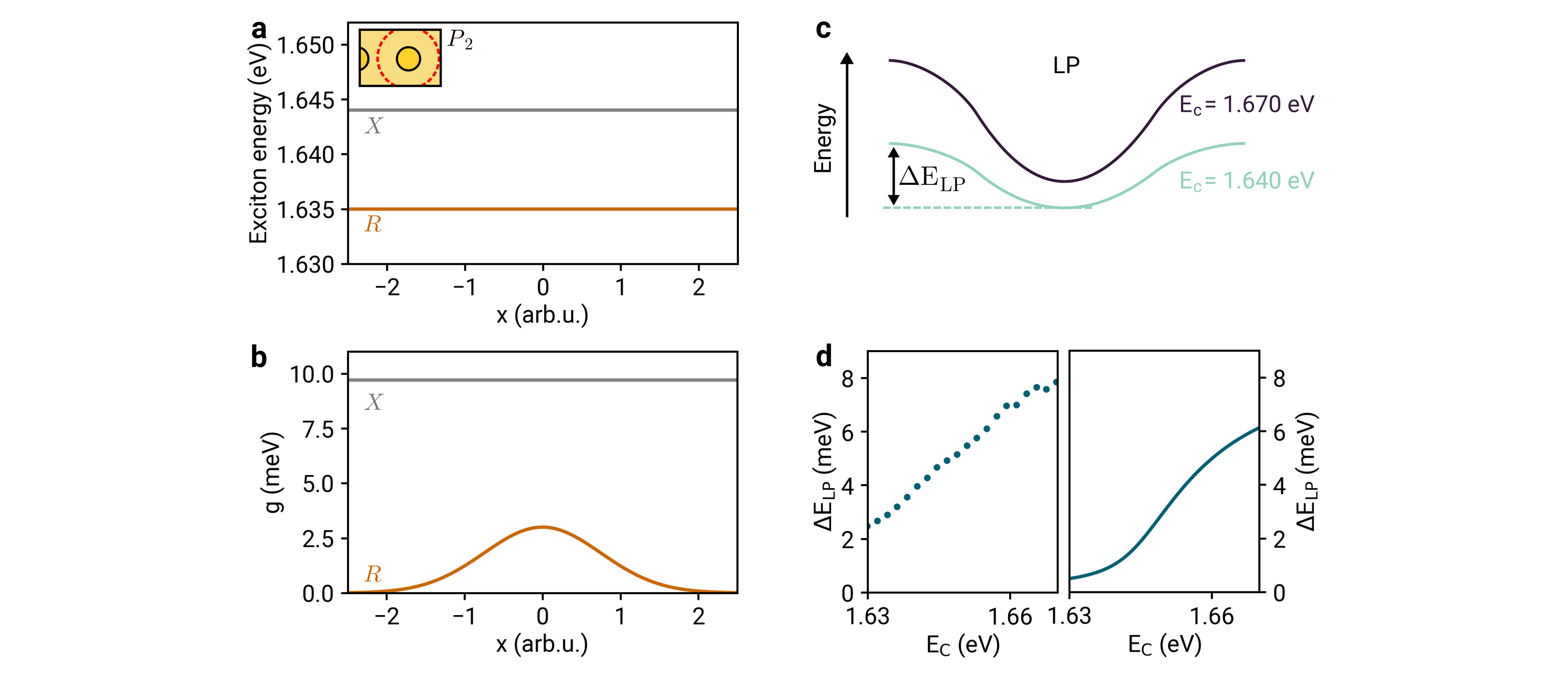}
	\caption{\textbf{Model for local polariton energy shifts.} \textbf{a}, \textbf{b}, Exciton energy and light-matter coupling strength, respectively, used to model the polariton energy shifts for the right etch site of P$_2$ illustrated in the inset. $X$ ($R$) resonances stem from excitons localized to the encapsulated monolayer (right etch site domain), as demonstrated in Supplementary Note~III. 
		\textbf{c},	Schematic of the lower polariton (LP) energy shift at a single exciton domain for different cavity energies, as observed in experiments presented in the main text. \textbf{d}, LP energy shift $\Delta E_\mathrm{LP}$ for the etch site $R$ of P$_2$ as obtained from experiment (left panel) and calculations as outlined in Supplementary Note~IV (right panel).} 
	\label{figS6}
\end{figure}

In Fig.~2 of the main text, we demonstrate local polariton energy shifts at the etch site positions. To visualize these shifts, we plot the transmission as a function of cavity mode position, measured at constant cavity energy $E_C$. $E_C$ was determined from a fit of the dissipative model of Supplementary Note~II at a position away from any etch site (e.g. $x = 0\ \mu$m). At each position, the spectrum was normalized to the maximum transmission of the lower polariton branch. The polariton energies, shown as dots in Fig.~2d of the main text, were determined as the energies of maximum cavity transmission of lower and upper polariton branch for each position. Polariton energies were fit with the double Gaussian profile
\begin{equation}
	E(x) = c - a_L\exp(-(x-x_L)^2/s_L^2) - a_R\exp(-(x-x_R)^2/s_R^2),
	\label{Eq:DGauss}
\end{equation}
where the fit results for $a_L$ and $a_R$ are the local energy shifts at the left and right etch site, respectively, shown Fig.~2c of the main text. For the single etch site in Fig.~2a of the main text, we modified Eq.~\ref{Eq:DGauss} to describe a single Gaussian well with the offset $c$ a linear function of position. 

To elucidate the mechanism of the local energy shift, we focus on the right etch site of P$_2$, as illustrated in the inset of Supplementary Fig.~\ref{figS6}a. Energy and light-matter coupling strength of the relevant exciton resonances are shown in Supplementary Figs.~\ref{figS6}a and b, respectively. The spatial dependence is based on the measurements presented in Supplementary Fig.~\ref{figS4}b, with the coupling strength for $R$ excitons approximated by a Gaussian profile with the etch site center positioned at $x = 0$ for illustration purposes. The blueshift in $X$ energy observed in Supplementary Fig.~4 is negligible for the computations presented in the following.

As evident from Fig.~2 of the main text, we observe an energy shift of magnitude $\Delta E_{\mathrm{LP}}$ for the lower polariton branch, which increases with increasing cavity energy, as illustrated in Supplementary Fig.~\ref{figS6}c. 
As explained in the main text, this observation results from the fact that as the cavity energy is increased, the lower polariton energy approaches that of the energetically lowest cavity-coupled exciton. As a result, $\Delta E_{\mathrm{LP}}$ monotonously increases until it reaches $E_X - E_R$. 

To provide additional verification, we compute the energy shift as $\Delta E_{\mathrm{LP}} =E_{LP}(x = \infty)-E_{LP}(x = 0)$, with the lower polariton energy $E_{LP}(x)$ obtained from the eigenstates of Eq.~\ref{Eq:H} for the values of $E$ and $g$ shown in Supplementary Figs.~\ref{figS6}a and b. The result, shown in the right panel of Supplementary Fig.~\ref{figS6}f, is in good agreement with the data obtained for P$_2$, which is reproduced in the left panel of the figure.

\section*{Supplementary Note V: Regime of dispersive cavity coupling}
\label{sec:SupSec4}

To describe our system in the regime of dispersive cavity coupling, i.e. for large detunings between cavity energy and all exciton resonances, we transform and expand the time-independent, non-dissipative Hamiltonian of Eq.~\ref{Eq:H} as described in the main text. We note that the resulting effective coupling of strength $J_{ij}$ has been studied for two or multiple individual (fermionic) two-level systems coupled to the same cavity mode, such as superconducting qubits~\cite{Majer2007} and defects in diamond~\cite{Evans2018}. This contrasts the case of bosonic excitons studied in our work. 

The strict theoretical requirement for Eq.~1 of the main text to hold is $\left|g_i^2/\Delta_i\right|^2 \ll 1$ for each excitonic species. In practice, we verified by numerical analysis that the deviation between the dissipative model and the true eigenstates of the full Hamiltonian of Eq.~\ref{Eq:H} depended on the precise values of coupling strengths and energies of the individual exciton resonances. For the data shown in Fig.~3 of the main text, two exciton resonances differing in energy by approximately 3~meV and with values of $g_i$ differing by a factor of approximately 6 coupled to the cavity with detunings $\Delta_X \approx 15$~meV. For these parameters, we found that deviations between polariton branches as predicted by Eq.~1 of the main text and the true eigenstates of the system were well within the typical linewidths observed in the experiments, underlining the validity of the approximation in the cases discussed in the main text. We note that this regime of dispersive cavity-detuning could only be accessed at selected positions on the sample. For different positions, broadening and reduced transmission of the polariton branches induced by disorder resulted in reduced signal-to-noise ratio, preventing the extraction of the system's eigenstates at satisfactory experimental confidence. 

The theoretical eigenstates of the effective Hamiltonian shown in Fig.~3b of the main text were computed for two cavity-coupled exciton resonances $X$ and $L$ using the following set of parameters: $E_X = 1.6436$~eV, $g_X = 9.5$~meV, $E_L = 1.6399$~eV, $g_L = 1.6$~meV. These values are in excellent agreement with the results obtained from fits of Eq.~\ref{Eq:T} to the transmission spectra for different cavity lengths measured at the same cavity mode position: $E_X = 1.64358 \pm 0.00002$~eV, $g_X = 9.577 \pm 0.004 $~meV, $E_L = 1.63999 \pm 0.00003$~eV, $g_L = 1.5 \pm 0.1$~meV. From the same fits, we obtained the cavity energy as a function of piezoelectric actuator voltage used in the theoretical computation.

The theoretical transmission profile shown in Fig.~4b of the main text was obtained from a fit of Eq.~2, yielding the parameters: $E_X = 1.6431 \pm 0.0002$~eV, $g_X = 9.5322 \pm 0.0001 $~meV, $E_L = 1.63750 \pm 0.00001$~eV, $g_L = 0.4 \pm 0.4$~meV, $E_R = 1.63582 \pm 0.00005$~eV, $g_R = 1.2 \pm 0.4$~meV. The values for $J_{LR}$ shown in Fig.~4c were computed from the measured values of $E_{L/R}$ and $g_{L/R}$ shown in Supplementary Fig.~4 for a cavity energy of 1.670~eV.
\clearpage
\begin{figure}
	\centering
	\includegraphics[scale=1]{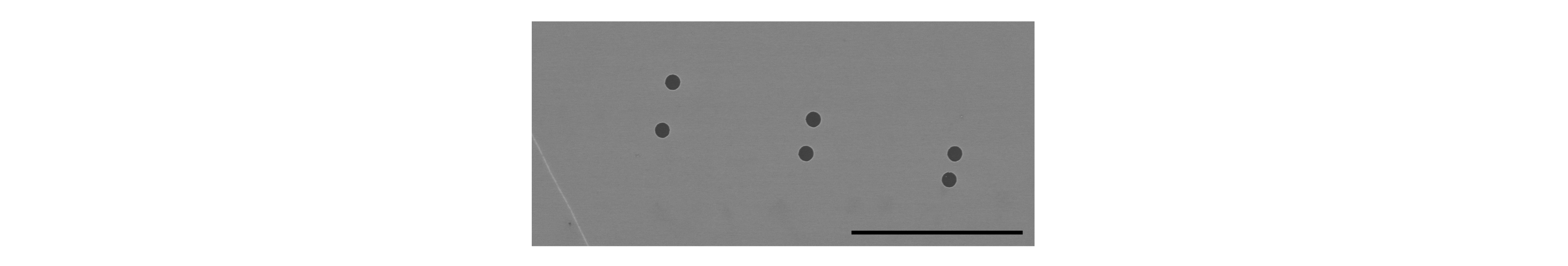}
	\caption{\textbf{Processing of hexagonal boron nitride.} SEM image of a processed hBN flake, showing pairs of circular air-holes also visible in Fig.~1b of the main text. An identical pattern was used for the flake in the fabricated device, which was not imaged in SEM to avoid contamination caused by carbon deposition. The scalebar is 10~$\mu$m.} 
	\label{figS7}
\end{figure}